\documentclass[number,citesort,dvips]{arxstspdf}
\usepackage{dcolumn}
\usepackage{graphicx}
\usepackage{flushend}
\usepackage{stfloats}
\volume{24}
\issue{4}
\pubyear{2009}
\firstpage{430}
\lastpage{450}
\doi{10.1214/09-STS311}

\makeatletter
\newcolumntype{d}[1]{D{.}{.}{#1}}
\makeatother

\begin{document}
\begin{frontmatter}

\title{A Bayesian Method for Detecting and Characterizing Allelic
Heterogeneity and Boosting Signals in Genome-Wide Association Studies}
\runtitle{Bayesian Method}

\begin{aug}
\author[a]{\fnms{Zhan} \snm{Su}},
\author[b]{\fnms{Niall} \snm{Cardin} \ead[label=u1,url]{http://www.wtccc.org.uk}},
\author[c]{\fnms{} \snm{the Wellcome Trust Case Control Consortium}},\\
\author[d]{\fnms{Peter} \snm{Donnelly}} \and
\author[e]{\fnms{Jonathan} \snm{Marchini}\ead[label=e4]{marchini@stats.ox.ac.uk}\corref{}}
\runauthor{Z. Su et al.}

\affiliation{University of Oxford}

\address[a]{Zhan Su is Analyst, Department of Statistics, University of Oxford, 1 South Parks
Road, Oxford OX1 3TG, UK and Wellcome Trust Centre for Human Genetics, University of Oxford,
Roosevelt Drive, Oxford, OX3 7BN, UK.}
\address[b]{Niall Cardin is Postdoctoral Recearcher, Department of Statistics, University of Oxford, 1 South Parks
Road, Oxford OX1 3TG, UK. Full details of Consortium membership are available at
\protect\href{http://www.wtccc.org.uk}{www.wtccc.org.uk}.}
\address[d]{Peter Donnelly is Professor, Wellcome Trust Centre for Human Genetics, University of Oxford,
Roosevelt Drive, Oxford, OX3 7BN, UK and Department of Statistics, University of Oxford, 1 South Parks
Road, Oxford OX1 3TG, UK.}
\address[e]{Jonathan Marchini is University Lecturer in Statistical Genomics, Department of Statistics, University of Oxford, 1 South Parks
Road, Oxford OX1 3TG, UK and Wellcome Trust Centre for Human Genetics, University of Oxford,
Roosevelt Drive, Oxford, OX3 7BN, UK \printead{e4}.}

\end{aug}

\begin{abstract}
The~standard paradigm for the analysis of genome-wide association
studies involves carrying out association tests at both typed and
imputed SNPs. These methods will not be optimal for detecting the signal
of association at SNPs that are not currently known or in regions where
allelic heterogeneity occurs. We propose a novel association test,
complementary to the SNP-based approaches, that attempts to extract
further signals of association by explicitly modeling and estimating
both unknown SNPs and allelic heterogeneity at a locus. At each site we
estimate the genealogy of the case-control sample by taking advantage of
the HapMap haplotypes across the genome. Allelic heterogeneity is
modeled by allowing more than one mutation on the branches of the
genealogy. Our use of Bayesian methods allows us to assess directly the
evidence for a causative SNP not well correlated with known SNPs and for
allelic heterogeneity at each locus. Using simulated data and real data
from the WTCCC project, we show that our method (i) produces a
significant boost in signal and accurately identifies the form of the
allelic heterogeneity in regions where it is known to exist, (ii) can
suggest new signals that are not found by testing typed or imputed SNPs
and (iii) can provide more accurate estimates of effect sizes in regions
of association.
\end{abstract}


\end{frontmatter}

\section{Introduction}

Over the last two years genome-wide association studies have been
successful in uncovering novel disease causing variants \cite{1,2,3,4,5,6,7}. All of
these studies have proceeded by testing for associations at SNPs assayed
by a commercial genotyping chip and many have also used genotype
imputation methods \cite{8} to test untyped SNPs, especially when combining
studies that used different genotyping chips to carry out larger
meta-analysis studies.

It is possible that signals of association will be missed by these
methods and there are several ways in which this could happen. First,
the true causal variant, which may be a SNP but could also be an Indel
or Copy Number Variant (CNV), may not be on the chip or on the typed
reference panel and may not be in sufficient Linkage Disequilibrium (LD)
with a single typed or imputed SNP for a signal to be detected. If this
is the case the variant may be well identified by considering a local
haplotype in the region, thus the association may be detected if such
effects are tested for association. Second, it may be the case that
the causal model of association in the region involves more than one
SNP. One way to describe this model would be to say that there is
allelic heterogeneity in the association signal. If the SNPs are in LD
then the various haplotypes that consist of the causal SNPs may have
distinct relative risks. If this is the case then the model might also
be described as a haplotype effect model.

In this paper we investigate a method that is complementary to SNP-based
association tests that allows for these more complex disease models. To
go beyond testing typed or imputed genetic variants we need to construct
a model for genetic variation that has not been directly observed. We
achieve this by modeling the genealogy of the sample of chromosomes at
each point along the genome and then estimating genotypes, in the
case-control samples, at SNPs derived by placing mutations on the
individual branches of the tree. The~genotypes that are derived from the
local genealogies can then be associated with the phenotype under study,
which we test using Bayesian methods that naturally account for the
inherent uncertainty in the location of the disease mutation on the
genealogy. Some previous approaches that have used genealogical trees,
have either been applicable only to haplotype data with no missing
data \cite{9,10} or computationally prohibitive and thus restricted to small
samples \cite{11,12}. The~method that we present here is applicable to genotype
data with missing data and is computationally feasible to analyze
thousands of individuals across the whole genome (it requires
approximately the same amount of computational resources as imputation
 \cite{8}). A novel feature of our method is that we can take advantage of the
HapMap haplotypes to build the genealogical trees at each putative risk
locus.

We provide an informal description of the method first and full
technical details of the method are given in the Methods section. Our
approach proceeds in several stages:

\begin{longlist}
\item[(i)] We use a \textit{panel} of known haplotype variation, such as
HapMap \cite{13}, and an estimate of the fine-scale recombination rate (such
as that available from the HapMap website), to construct an
approximation to the genealogy of the sample of haplotypes in the panel,
at each point on a grid of positions across the genome.

\item[(ii)] For each such tree, we then, in turn, consider putative
mutations on each of its branches. Once the branch for a mutation is
chosen, this will fix the alleles carried by chromosomes at each of the
tips of the tree. Assuming such a SNP exists in the population, we use a
population genetics model to predict the likely genotypes at this SNP in
each case and control individual (we call these the \textit{study}
individuals). This is perhaps simplest to conceptualize under the
simplifying assumption that we had haploid data on the study
individuals. The~population genetics model allows us to place each
haploid study chromosome (probabilistically) on the tips of the
genealogical tree: each tip contains a single panel haplotype, and the
study haplotype will tend to be placed on the tips corresponding to
panel haplotypes that are locally similar to it. For diploid study data
there is an additional level that, in effect, averages over likely local
phasings of the data. The~result, for each study individual, is a
probability distribution over the possible genotypes at the putative
SNP.

\item[(iii)] The~next step takes the predicted genotypes with their
uncertainties and looks for evidence of association with disease status.
(Note the need to handle appropriately the uncertainty over the
predicted genotypes.) We do this here in a Bayesian framework. For a
particular genomic location and putative SNP, the evidence for
association is naturally measured by a Bayes factor \cite{8} (BF) that
compares a model of association with a null model of no association. The~uncertainty over the possible branch carrying the mutation is handled by
averaging over the Bayes factors for each branch, to give a single BF
summarizing the evidence for the presence of a causative mutation at
that position.

\item[(iv)] To allow for possible allelic heterogeneity, we can extend
the analyses above by putting two (or more) putative mutations on the
genealogical tree and predicting genotypes for the pair of SNPs in study
individuals, before fitting disease models with multiple causative
mutations. We would then average over pairs of positions for the
mutation in calculating a BF for the strength of evidence for a
\mbox{2-mutation} disease model at that position, compared to the null
hypothesis, and by comparing the \mbox{2-mutation} BF with the single mutation
BF, one can assess the relative evidence for allelic heterogeneity, for
example.

\item[(v)] At genomic positions where there is a signal of association,
we can combine the estimated tree with the most likely mutation pattern
to characterize graphically the signal of association, and identify
which local haplotypes show evidence of differential disease risk
between cases and controls.
\end{longlist}

We have applied our method to data from the Wellcome Trust Case Control
Consortium (WTCCC) \cite{5} to assess its performance and illustrate its novel
features. Specifically, we have applied it to several risk loci that are
known to exhibit allelic heterogeneity (e.g., the \textit{NOD2}
region for Crohn's disease) and show that our method provides a boost in
signal over testing both typed and imputed SNPs. In addition we show
that the method can accurately identify the branches on the genealogy
that correspond to the true causal variants. Further we have applied the
method across the genome for all seven diseases studied by WTCCC and
have compared its performance to testing both typed and imputed SNPs.
Our method is able to identify (subsequently validated) associations not
picked up by tested typed or imputed SNPs and results in a much richer
characterization of the associated signal in several regions. We show
that no one method is optimal in detecting association but that the new
method presented here clearly have a role to play in detecting and
characterizing associations in genome-wide scans. Our use of Bayesian
methods allows the Bayes factors at both typed and imputed SNPs to be
naturally combined with the Bayes factors produced by our method.

We also carried out some simulation studies that highlight additional
features of our approach. First, we examine how well our method does
at uncovering allelic heterogeneity where it exists. We show that this
can be quite a hard problem but our method does have good power to
uncover the action of more than one causal variant. Second, we
consider the problem of estimating the effect size of a causal variant
in an associated region in the specific case where the causal variant is
not well tagged by a typed SNP. We show that our method is able to
provide a more accurate estimate of the effect size in this case.

The~next section describes the details of our methods and this is
followed by a section on the analysis of the WTCCC data and simulation
studies. We conclude with a discussion of the results and the likely
applications of our method.

\section{Methods}

We use $H_{i} = \{ H_{1},\ldots ,H_{N}\}$ to denote a set of $N$ known
haplotypes, where $H_{i} = (H_{i1},\ldots ,H_{iL})$ is a single
haplotype, $H_{ij} \in\{ 0,1\}$ and $L$ is the number of SNP loci. For
all the analysis in this paper we have set $H$ to be the 120 CEU
haplotypes estimated as part of the HapMap project \cite{13}. We let $G = \{
G_{1},\ldots ,G_{K}\}$ denote the genotype data for the $K$ individuals in a
new study, where $G_{i} = (G_{i1},\ldots ,G_{iL})$ and $G_{ij} \in\{
0,1,2,\mathrm{missing}\}$. It is likely that many of the genotypes will
be missing since genome-wide SNP chips do not contain every SNP in the
HapMap panel. We use $\Phi_{i} \in\{ 0 = \mathrm{Control},1 =
\mathrm{Case} \}$ to denote the binary phenotype of the $i$th
individual. Let $X = \{ X_{1}, \ldots,X_{M} \}$ be a grid of physical
positions for carrying out association tests; for our analysis, we use a
grid spacing of 5~kb on every chromosome.

\textit{Step 1}. A genealogical tree, $T$, is constructed at every
position in $X$ using the set of known haplotypes. The~trees are built
using the coalescent model with recombination and approximate the
posterior modal tree given the haplotypes. To do this it is useful to be
able consider $P(T|H)$ under the coalescent. Using the Bayes Formula we can
rewrite this as: $P(H|T)P(T)/P(H)$. Although it is simple to calculate
these values under the coalescent with simple mutation models it is not
known how to simulate directly from this distribution, or how to produce
trees that maximize this expression \cite{14}.

To make this task simpler it is helpful to factorize this expression
into the individual events that make up the tree (coalescence,
recombination or mutation). It is useful to note that trees augmented
with mutation track the haplotypes backward in time, and these
haplotypes change after each event. Note that $P(T) = \prod_{i}
P(E_{i})$ where $i$ indexes the events backwards in time and $E_{i}$ is
the $i$th event. Also $P(T|H) = \prod_{i} P(E_{i}|H_{i})$, where
$H_{i}$ denotes the haplotypes as changed by the first $i$ events. Then,
note that $P(E_{i}|H_{i}) = P(H_{i}|E_{i})P(E_{i})/P(H_{i})$.

It is not known how to calculate $P(H_{i})$ directly. However, as the
coalescent is Markov backward in time $P(H_{i}|E_{i})$ (the probability
of the haplotypes $H_{i}$ given that the next event backwards in time
is $E_{i})$ is equal to $P(H_{i + 1})$ (the probability of the haplotypes
as changed by the event $E_{i})$. So to calculate $P(E_{i}|H_{i})$ it is
only necessary to calculate $(P(H_{i + 1})/P(H_{i}))P(E_{i})$. For all
types of event (coalescence, recombination or mutation) the quotients
$P(H_{i + 1})/P(H_{i})$ simplify to give terms of the form $P(H_{n +
1}|H_{1},\ldots ,H_{n})$. These terms still cannot be calculated efficiently
under the coalescent, however they are amenable to approximation using
Hidden Markov Models \cite{15,16}.

Once these values can be approximated it is possible to generate a tree
that approximates the modal posterior tree as follows:

\begin{enumerate}
\item[1.] Initialize: Decide on mutation model, recombination rates, and
initialize the haplotypes, $H_{0}$, as the set of known haplotypes input
to the method.

\item[2.] Recursion (steps 2 through 6): Enumerate all possible events
that may be the next event backwards in time.

\item[3.] For each of these events approximate $P(E_{i}|H_{i})$, the
posterior probability of each event, as described above.

\item[4.] Choose the event with the highest posterior probability.

\item[5.] Generate haplotypes $H_{i + 1}$ by applying the chosen event
to haplotypes $H_{i}$.

\item[6.] Stop: When each locus has reached its common ancestor the
process terminates.
\end{enumerate}

We used the recombination rates estimated from the HapMap \cite{13}, and an
infinite sites mutation model for this analysis. This step needs only be
performed once for each set of reference haplotypes. For example, we
have calculated and stored a set of trees for the CEU HapMap haplotypes
across the genome at a grid of positions with a 5~kb spacing between
positions. Trees produced by this method (called TREESIM) may be useful
for other population genetics inferences.

\textit{Step 2}. Given the genealogical tree at a given position,
$X_{m}$, estimated in step 1 our method works by averaging over
locations of the disease causing mutations on branches, $b$, of the
tree. Each mutation defines a hypothetical disease SNP that can be added
into the panel of haplotypes. For each individual we use a model to
calculate the expected allele count for this disease mutation at the
position $X_{m}$. We use $H^{mb}$ to denote the set of haplotypes, $H$,
augmented with the disease SNP at the position $X_{m}$ created by a
mutation on branch $b$\vspace*{2pt} and $G_{i}^{mb}$ to denote the genotype vector for
study individual $i$ augmented with the (unknown) genotype for the
branch $b$ disease SNP at position $X_m$. We use a model similar to that used in
IMPUTE \cite{8} that relates each individual's genotype vector to the set of
known haplotypes, $P( G_{i}^{mb}| H^{mb})$, as a Hidden Markov Model
in which the hidden states are a sequence of pairs of the $N$ known
haplotypes in the set $H$. That is,
\begin{eqnarray*}
P( G_{i}^{mb}| H^{mb} ) &=& \sum_{Z_{i}^{(1)},Z_{i}^{(2)}} P\bigl(
G_{i}^{mb}| Z_{i}^{(1)},Z_{i}^{(2)},H^{mb} \bigr)\\
&&\hspace*{32pt}{}\cdot  P\bigl(
Z_{i}^{(1)},Z_{i}^{(2)}| H^{mb} \bigr),
\end{eqnarray*}
where $Z_{i}^{(1)} = \{ Z_{i1}^{(1)},\ldots ,Z_{id}^{(1)},\ldots ,Z_{i(L +
1)}^{(1)} \}$ and $Z_{i}^{(2)} = \{
Z_{i1}^{(2)},\ldots ,Z_{id}^{(2)},\ldots ,Z_{i(L + 1)}^{(2)} \}$ are the two
sequences of hidden states at the $L + 1$ sites, $Z_{il}^{(j)} \in\{
Z_{i1}^{(2)},\ldots ,\break Z_{id}^{(2)},\ldots ,Z_{i(L + 1)}^{(2)} \}$, and $d$ is the
position of the disease SNP in the augmented sets $H^{mb}$
and $G_{i}^{mb}$. These hidden states can be thought of as the pair of
haplotypes in the set $H$ that are being copied to form the genotype
vector $G_{i}^{mb}$. The~term $P( Z_{i}^{(1)},Z_{i}^{(2)}| H^{mb} )$
defines our prior probability on how sequences of hidden states change
along the sequence and\break $P( G_{i}^{mb}| Z_{i}^{(1)}, Z_{i}^{(2)},H^{mb})$ models how the observed genotypes will be close to but not exactly
the same as the haplotypes being copied.

The~expected genotype at the disease SNP can be defined as
\begin{eqnarray*}
e_{i}^{mb} &=& \mathrm{E}( G_{im}^{mb} )\\
 &=& \sum_{k_{1} = 1}^{N}
\sum_{k_{2} = 1}^{N} \bigl( I( H_{k_{1}m}^{mb} = 1 ) + I( H_{k_{2}m}^{mb} = 1
) \bigr)\\
&&\hspace*{37pt}{}\cdot p_{im}( k_{1},k_{2} ),
\end{eqnarray*}
where $I$ is the indicator function and
\begin{eqnarray*}
&&p_{im}( k_{1},k_{2} )\\
&&\quad  = P\bigl( \bigl\{
Z_{im}^{(1)},Z_{im}^{(2)} \bigr\} = \{ k_{1},k_{2} \}|
G_{i}^{mb},H^{mb} \bigr)\\
&&\quad  \propto P\bigl( G_{i}^{mb}|
\bigl\{ Z_{im}^{(1)},Z_{im}^{(2)} \bigr\} = \{ k_{1},k_{2}
\},H^{mb} \bigr)\\
&&\quad  =\mathop{\sum_{Z_{i}^{(1)},Z_{i}^{(2)}:}}_{\{Z_{im}^{(1)},Z_{im}^{(2)} \}= \{ k_{1},k_{2} \}}
P\bigl( G_{i}^{mb}| Z_{i}^{(1)},Z_{i}^{(2)},H^{mb} \bigr)\\
&&\hspace*{90pt} {}\cdot P\bigl( Z_{i}^{(1)},Z_{i}^{(2)}| H^{mb} \bigr).
\end{eqnarray*}

This step involves a calculation that is practically identical to that
used in the method IMPUTE, which has been used in several genome-wide
analyses to date, and illustrates that the method is practical for this
type of analysis.

\textit{Step 3}. The~final step involves evaluating whether there is
evidence of association at each position by calculating a BF between a
model of association $M_{1}$ and a model of no association $M_{0}$. The~simplest way of modeling association at the disease SNP created by
placing a mutation on branch $b$ at position $X_{m}$ is to create a $2\times
2$
table of expected allele counts\vspace*{6pt}
\noindent\begin{tabular*}{\columnwidth}{@{\extracolsep{\fill}}lcc@{}}
\hline
 & \textbf{0} & \textbf{1}\\
 \hline
Controls & $n_{00} = n_{U} - n_{01}$ & $n_{01} = \sum_{i:\Phi _{i} = 0} e_{i}^{mb}$\\
Cases & $n_{10} = n_{A} - n_{11}$ & $n_{11} = \sum_{i:\Phi _{i} = 1} e_{i}^{mb}$\\
\hline
\end{tabular*}

\noindent where $n_{U}$ and $n_{A}$ are the numbers of unaffected (control) and
affected (case) haplotypes respectively.

From this table we can calculate a Bayes factor as $\mathit{BF}_{mb} = \frac{P(
\mathit{Data}|M_{1} )}{P( \mathit{Data}|M_{0} )}$, where
\begin{eqnarray*}
&&\hspace*{-4pt} P( \mathit{Data}| M_{1}
)\\
&&\hspace*{-4pt}\quad  = \int P( \Phi | e^{mb},\theta _{1};M_{1} ) P( \theta _{1}| M_{1})\,d\theta _{1}\\
&&\hspace*{-4pt}\quad  = \int p^{n_{11}}( 1 - p )^{n_{01}}\frac{\Gamma ( a +
c )}{\Gamma ( a )\Gamma ( c )} p^{a - 1}( 1 - p )^{c - 1}\,dp\\
&&\hspace*{-4pt}\qquad {}\cdot \int
q^{n_{10}}( 1 - q )^{n_{00}}\frac{\Gamma ( a + c )}{\Gamma ( a )\Gamma (
c )} q^{a - 1}( 1 - q )^{c - 1}\,dq\\
&&\hspace*{-4pt}\quad  = \frac{\Gamma ( n_{11} + a
)\Gamma ( n_{01} + c )}{\Gamma ( n_{0} + a + c )}\\
&&\hspace*{-4pt}\qquad {}\cdot\frac{\Gamma ( n_{10} +
a )\Gamma ( n_{00} + c )}{\Gamma ( n_{1} + a + c )}\\
&&\hspace*{-4pt}\qquad {}\cdot \biggl[ \frac{\Gamma ( a +
c )}{\Gamma ( a )\Gamma ( c )} \biggr]^{2},
\end{eqnarray*}
where $p$ and $q$ are penetrance parameters of the alleles~1 and 0
respectively, and
\begin{eqnarray*}
&&P( \mathit{Data}| M_{0} )\\
&&\quad  = \int P(
\Phi | e^{mb},\theta _{0};M_{0} ) P( \theta
_{0}| M_{0} )\,d\theta _{0}\\
&&\quad  = \int r^{n_{A}}(
1 - r )^{n_{U}}\frac{\Gamma ( a + c )}{\Gamma ( a
)\Gamma ( c )} r^{a - 1}( 1 - r )^{c - 1}\,dr\\
&&\quad  = \frac{\Gamma ( n_{A} + a )\Gamma ( n_{U} + c
)}{\Gamma ( n_{A} + n_{U} + a + c )}\frac{\Gamma (
a + c )}{\Gamma ( a )\Gamma ( c )},
\end{eqnarray*}
where $r$ is a penetrance parameter unconditional on allele. These
calculations utilize a Binomial likelihood for the expected allele
counts and a $\operatorname{Beta}(a,c)$ prior on the parameters of the model.
For the analysis of the WTCCC data in this paper we used a
$\operatorname{Beta}(20,30)$ prior the parameters $p, q$ and $r$ in the models.
This prior is centered on the proportion of cases and controls in the
sample and leads to a distribution on the relative risk ($p/q$)
with mean 1.0 and standard deviation of 0.49. Supplementary Figure 3
illustrates the prior on the relative risk.

The~additive model of association we have used is the simplest option
and was chosen initially for computational convenience. One criticism of
this model is that it implicitly makes an assumption of Hardy--Weinberg
Equilibrium (HWE) at the SNP in both cases and controls and is more
susceptible to the effects of population structure \cite{17}. To ameliorate
these concerns we have also developed a facility to output estimated (or
imputed) genotypes, in the case-control samples, at SNPs derived by
placing mutations on the individual branches of the tree. These SNP
genotypes can be fed directly into our software SNPTEST thus allowing a
range of more sophisticated models to be applied to the data, such as
standard additive, dominant, recessive and general tests of association
and tests that condition on covariates and testing of other more refined
phenotypes. This facility is used in one of our simulation studies where
we investigate the performance of our method in estimating effect sizes
in associated regions.

A Bayes factor for the position $X_{m}$ can be obtained by averaging the
Bayes factors for each branch, $b$, weighted by the prior, $P(b)$, on
each branch that was estimated in step 1:
\[
\mathit{BF}_{m} = \sum_{b \in B} \mathit{BF}_{mb} P(b).
\]

We take $P(b)$, the prior probability of a mutation occurring on a branch $b$, to be proportional to the expected length of $b$ under the
coalescent, that is, $\sum_{i = m}^{n} \frac{2.0}{i(i - 1)}$, where $
m $ and $ n$ are the number of distinct branches when $b
$ was first formed and just before $m$ coalesced
respectively. Our prior favors mutations that occur on long branches.

An analogous set of calculations can be carried out by assuming that
there exist two (or more) distinct disease mutations on branches of the
tree at each position. The~prior probability of mutations on more than
one branch is simply the product of the probabilities of mutation
occurring on each individual branch.

\begin{figure*}[b]

\includegraphics{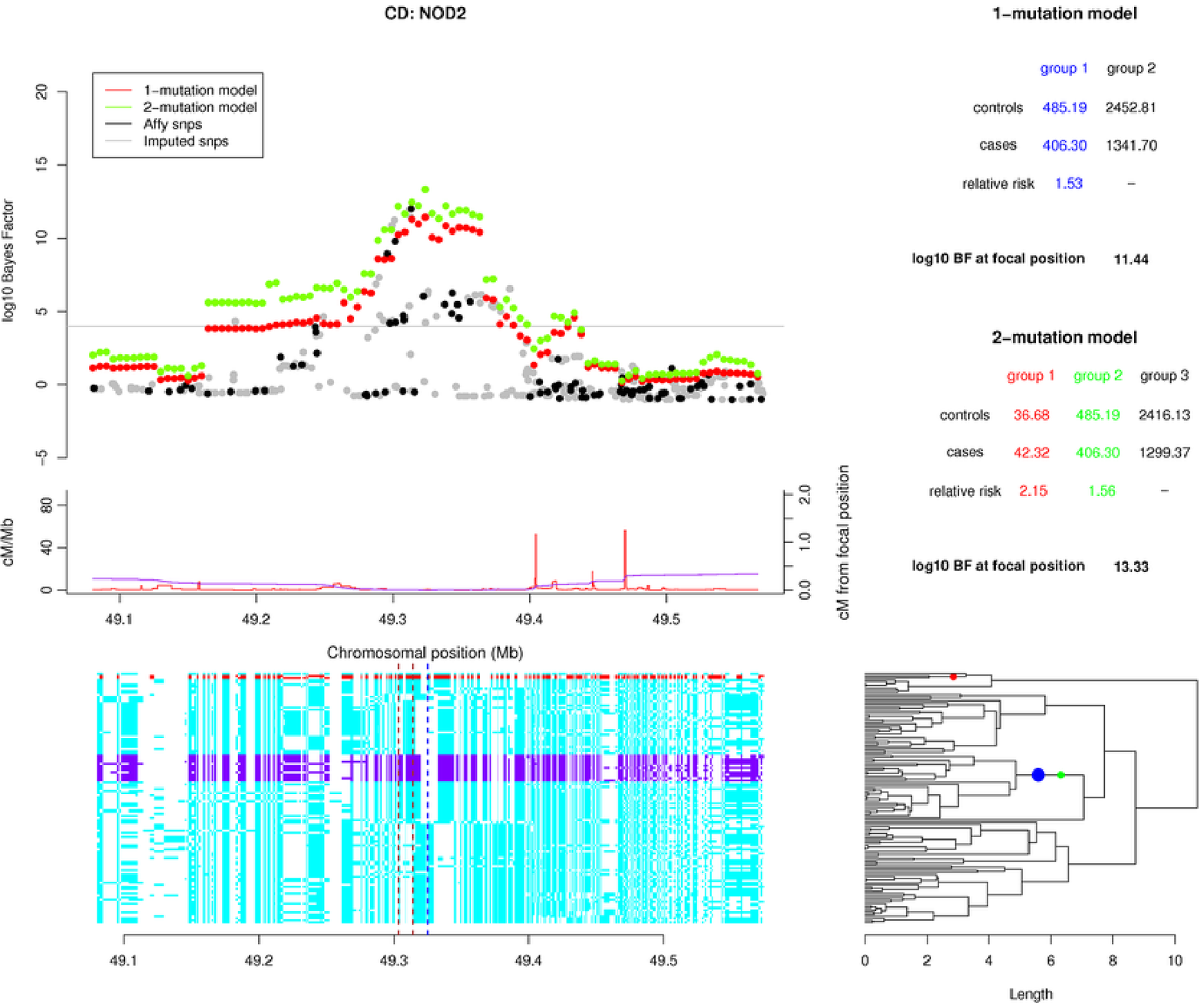}

\caption{The~top left panel of the plot shows the $\log_{10}$ Bayes
factor for the \mbox{1-mutation} model (red) and \mbox{2-mutation} model (green)
within the \textit{NOD2} region of the Crohn's Disease analysis. The~recombination
map (red line) and the cumulative recombination map
(purple line) are shown below this. The~bottom left panel shows the 120
CEU HapMap haplotypes across the region. Each row of this panel is a
haplotype and each column is a SNP. The~haplotypes are colored to
indicate the three haplotypes that occur at the 2 coding SNPs rs2066844
and rs2066845 (red${}={}$CC, purple${}={}$TG, cyan${}={}$CG). The~dashed vertical blue
and brown lines indicate the position of the largest $\log_{10}$ Bayes
factor for the \mbox{2-mutation} model (the focal position) and the two coding
SNPs respectively. The~bottom right panel shows the estimated
genealogical tree at the focal position. The~$x$-axis of the plot was
chosen to provide a clear view of all the branches in the tree. The~branches associated
with the best \mbox{1-mutation} and \mbox{2-mutation} models that
make the largest contributions to the Bayes factors are shown with blue
and red/green dots respectively. The~top right panel shows the tables of
expected allele counts for the \mbox{1-mutation} and \mbox{2-mutation} models together
with a summary of the Bayes factors that occur at the focal position.
The~columns of the tables are color matched to the mutations on the tree
in the bottom right panel.}\label{fig1}
\end{figure*}

\subsection{Posterior Probability of the Number of~Mutations}

Let $\mathit{BF}_{1}$ and $\mathit{BF}_{2}$ be the Bayes factor under the \mbox{1-mutation} model
($M_{1}$) and \mbox{2-mutation} model ($M_{2}$) respectively, then the Bayes
factor, $\mathit{BF}$, comparing the \mbox{2-mutation} model to the \mbox{1-mutation} model is
given by
\[
\mathit{BF} = \frac{P(D|M_{2})}{P(D|M_{1})} = \frac{P(D|M_{2}) /
P(D|M_{0})}{P(D|M_{1}) / P(D|M_{0})} = \frac{\mathit{BF}_{2}}{\mathit{BF}_{1}},
\]
 where $D$
is data and $M_{0}$ is the null model. If we assume a prior odds for two
mutations vs. one mutation of $1\dvtx 1$ then the posterior odds is simply $\mathit{BF}$,
the ratio of $\mathit{BF}_{2}$ and $\mathit{BF}_{1}$.

\section{Results}

\subsection{Application to \textit{NOD2} Locus}

To illustrate the utility of our method on an established disease locus
exhibiting allelic heterogeneity we applied our approach to the
\textit{NOD2} locus \cite{18,19} for Crohn's disease on chromosome 16. We
applied our approach to this region using a set of trees built using the
CEU HapMap haplotypes at 5~kb intervals throughout the region. We used,
after filtering, 1748 case and 2938 control individuals genotyped as
part of the WTCCC study.

The~results produced by our method are shown in Figure \ref{fig1}, which compares
the signals of association at SNPs on the Affymetrix chip, imputed SNPs
and two versions of our method that allow one and two mutations on the
tree at each position respectively. All the methods show a substantial
signal at the locus but the signal for our new methods are higher and
broader. The~signals are also much smoother across the region than the
signals from the typed and imputed SNPs. The~$\log_{10}$ Bayes factors
allowing for one and two mutations peak at 11.44 and 13.33 respectively
(larger values of the Bayes factor indicate stronger evidence for
association). These compare favorably with the $\log_{10}$ Bayes factors
at the best Affymetrix SNP (12.00) and the best imputed SNP (11.42); so
the new method provides a stronger signal than comparable current
approaches.

Next we can assess the relative evidence for the \mbox{2-mutation} model
compared with the \mbox{1-mutation} model simply by dividing the relative Bayes
factors, or equivalently through the difference of the $\log_{10}$ Bayes
factors. Here the latter is 1.89, indicating that the data is about
$10^{1.89} = 78$ times more likely under the \mbox{2-mutation} model than the
\mbox{1-mutation} model. If the \mbox{1-~and 2-mutation} models were thought equally
likely \textit{a priori} this would imply a posterior probability of
0.987 for two mutations versus one mutation indicating substantial
evidence of allelic heterogeneity.

\begin{table*}
\tabcolsep=0pt
\caption{Regions that exhibited a $\log_{10}$ Bayes factor greater than 4
for either the \mbox{1-mutation} or the \mbox{2-mutation} model in the analysis of the
seven WTCCC diseases. \textup{Log}$_{10}$ Bayes factors for \mbox{1-mutation} and
\mbox{2-mutation} models are given together with the posterior probability of
the \mbox{2-mutation} model relative to the \mbox{1-mutation} model. \textup{Log}$_{10}$ Bayes
factors and $p$-values are also given for the~best~Affymetrix~SNP~and~best~imputed SNP in the regions}
\label{tab1}
\begin{tabular*}{\textwidth}{@{\extracolsep{4in minus 4in}}ld{2.0}cd{3.2}d{3.2}d{1.2}d{3.2}d{3.2}d{1.8}d{1.8}@{}}
\hline
 &  &  & \multicolumn{1}{c}{\textbf{1-mutation}} & \multicolumn{1}{c}{\textbf{2-mutation}} & \multicolumn{1}{c}{\textbf{Prob.}} & \multicolumn{1}{c}{\textbf{Affy.}}  & \multicolumn{1}{c}{\textbf{IMPUTE}}  & \multicolumn{1}{c}{\textbf{Affy.}} & \multicolumn{1}{c@{}}{\textbf{IMPUTE}} \\
\multicolumn{1}{@{}c}{\textbf{Disease}} & \multicolumn{1}{c}{\textbf{Chr.}} &\multicolumn{1}{c}{\textbf{Region (Mb)}}  & \multicolumn{1}{c}{\textbf{Log}$_{\bolds{10}}$
\textbf{BF}}
& \multicolumn{1}{c}{\textbf{Log}$_{\bolds{10}}$ \textbf{BF}} & \multicolumn{1}{c}{\textbf{2 mut.}} & \multicolumn{1}{c}{\textbf{Log}$_{\bolds{10}}$ \textbf{BF}} &
\multicolumn{1}{c}{\textbf{Log}$_{\bolds{10}}$ \textbf{BF}} & \multicolumn{1}{c}{$\bolds{p}$\textbf{-value}} & \multicolumn{1}{c@{}}{$\bolds{p}$\textbf{-value}}\\
 \hline
CAD & 9 & 21.98--22.11 & 11.04 & 11.01 & 0.48 & 11.66 & 11.58 & 1.79\mathrm{e}{-}14 & 1.48\mathrm{e}{-}14\\
CD & 1 & 67.25--67.47 & 12.96 & 17.99 & 1.00 & 10.07 & 15.82 & 6.45\mathrm{e}{-}13 & 7.93\mathrm{e}{-}18\\
CD & 2 & 233.93--233.99 & 10.38 & 10.29 & 0.45 & 11.11 & 11.55 & 7.10\mathrm{e}{-}14 & 2.79\mathrm{e}{-}14\\
CD & 5 & 40.33--40.65 & 10.45 & 14.68 & 1.00 & 10.41 & 10.93 & 2.13\mathrm{e}{-}13 & 1.32\mathrm{e}{-}13\\
CD & 5 & 131.65--131.83 & 5.82 & 6.13 & 0.67 & 4.54 & 7.18 & 5.40\mathrm{e}{-}07 & 3.04\mathrm{e}{-}10\\
CD & 5 & 150.16--150.3\phantom{0} & 4.94 & 4.89 & 0.47 & 5.43 & 5.51 & 4.26\mathrm{e}{-}08 & 3.15\mathrm{e}{-}08\\
CD & 6 & 31.36--31.39 & 4.61 & 4.69 & 0.55 & 1.96 & 6.52 & 0.000254 & 5.63\mathrm{e}{-}08\\
CD & 6 & 31.99--32.52 & 4 & 4.75 & 0.85 & 1.4 & 3.33 & 0.00106 & 7.13\mathrm{e}{-}07\\
CD & 10 & 101.27--101.29 & 5.32 & 5.4 & 0.55 & 5.91 & 6.05 & 1.41\mathrm{e}{-}08 & 1.03\mathrm{e}{-}08\\
CD & 16 & 49.16--49.44 & 11.44 & 13.33 & 0.99 & 12 & 11.42 & 5.78\mathrm{e}{-}15 & 7.20\mathrm{e}{-}17\\
CD & 18 & 12.77--12.87 & 5.56 & 5.52 & 0.48 & 5.42 & 5.53 & 4.56\mathrm{e}{-}08 & 1.72\mathrm{e}{-}09\\
RA & 1 & 113.59--114.26 & 20.73 & 20.81 & 0.55 & 22.36 & 11.87 & 4.90\mathrm{e}{-}26 & 3.92\mathrm{e}{-}18\\
RA & 6 & 29.66--33.77 & 102.92 & 124.67 & 1.00 & 74.84 & 91.19 & 3.44\mathrm{e}{-}76 & 1.98\mathrm{e}{-}106\\
T1D & 1 & 113.59--114.23 & 20.76 & 20.7 & 0.47 & 23.07 & 13.21 & 1.17\mathrm{e}{-}26 & 1.98\mathrm{e}{-}18\\
T1D & 4 & 123.59--123.88 & 4.59 & 4.58 & 0.49 & 4.42 & 5.63 & 5.00\mathrm{e}{-}07 & 2.24\mathrm{e}{-}07\\
T1D & 6 & 25.98--33.93 & 290.18 & {>}300 & 1.00 & 306.95 & 202.71 & 1.02\mathrm{e}{-}287 & 2.28\mathrm{e}{-}204\\
T1D & 10 & 6.13--6.15 & 4.35 & 4.85 & 0.76 & 3.31 & 4.58 & 7.97\mathrm{e}{-}06 & 3.19\mathrm{e}{-}07\\
T1D & 12 & 54.66--54.78 & 7.65 & 7.72 & 0.54 & 8.89 & 8.02 & 1.14\mathrm{e}{-}11 & 2.30\mathrm{e}{-}11\\
T1D & 12 & 109.83--111.48 & 10.98 & 10.98 & 0.50 & 12.53 & 12.74 & 2.17\mathrm{e}{-}15 & 2.06\mathrm{e}{-}16\\
T1D & 15 & 58.57--58.58 & 3.2 & 4.06 & 0.88 & 1.08 & 1.98 & 0.00242 & 4.46\mathrm{e}{-}05\\
T1D & 16 & 10.97--11.12 & 5.2 & 5.24 & 0.52 & 5.76 & 6.27 & 2.22\mathrm{e}{-}08 & 8.50\mathrm{e}{-}09\\
T2D & 6 & 20.79--20.81 & 4.04 & 4.15 & 0.56 & 4.15 & 4.35 & 1.02\mathrm{e}{-}06 & 1.01\mathrm{e}{-}07\\
T2D & 9 & 22.12 & 5.61 & 5.71 & 0.56 & 1.53 & 2.90 & 0.000706 & 2.22\mathrm{e}{-}05\\
T2D & 10 & 114.72--114.81 & 9.73 & 9.89 & 0.59 & 10.14 & 11.09 & 5.68\mathrm{e}{-}13 & 6.08\mathrm{e}{-}14\\
T2D & 16 & 52.36--52.38 & 5.01 & 5.11 & 0.56 & 5.89 & 5.74 & 1.44\mathrm{e}{-}08 & 2.07\mathrm{e}{-}08\\
\hline
\end{tabular*}
\end{table*}

There are three known coding SNPs in this region \cite{18,19}. Two of these
SNPs (rs2066845 and rs2066844) are in the HapMap panel. Figure \ref{fig1} shows
that the three distinct haplotypes induced by these two SNPs correspond
well to those identified by the best fitting \mbox{2-mutation} model. For
example, one of the two best mutations (red) precisely identifies the
CEU haplotypes that carry the rare rs2066845 mutation while the other
mutation (green) is only one branch away from precisely identifying the
haplotypes that carry the more common rs2066844 mutation. In other
words, our analyses of the WTCCC data using the new method go very close
to recovering the known pattern of disease susceptibility, based on much
more extensive genotyping. Relative risk estimates of red and green
mutations on the tree, relative to a lack of either of these mutations,
are 2.15 and 1.56 respectively.

\subsection{Application to the WTCCC Data}

We have applied this method to all seven genome-wide association studies
carried out as part of the WTCCC study \cite{5}. Doing this allows us to
compare the performance of the new method to those methods that are
currently routinely used to analyze genome-wide association studies,
that is, analysis of genotyped SNPs on the chip and of imputed SNPs.

Table \ref{tab1} lists the regions that exhibited a $\log_{10}$ Bayes factor
greater than 4 for either the \mbox{1-mutation} or the \mbox{2-mutation} model. Just
as with $p$-values, there is no \textit{correct} threshold for Bayes
factors for ``declaring'' association. Several arguments suggest that
the threshold on which we focus here is quite a stringent one.
Empirically, many SNPs with lower single-SNP Bayes factors in the WTCCC
data are now known to correspond to real effects, and most or all SNPs
meeting this threshold have been replicated. On a theoretical level,
this is the required threshold in order for the posterior odds of
association at a site to be greater than 1 when using a prior odds of
association of 1$/$10,000. This prior is motivated by the argument that
there are on the order of 1,000,000 ``independent'' regions of the
genome and an expectation of 100 of these being involved in the disease.
Most of the regions in this table were identified by the SNP and
imputation analysis of the main WTCCC study but there are some notable
differences.

\begin{figure*}[t]

\includegraphics{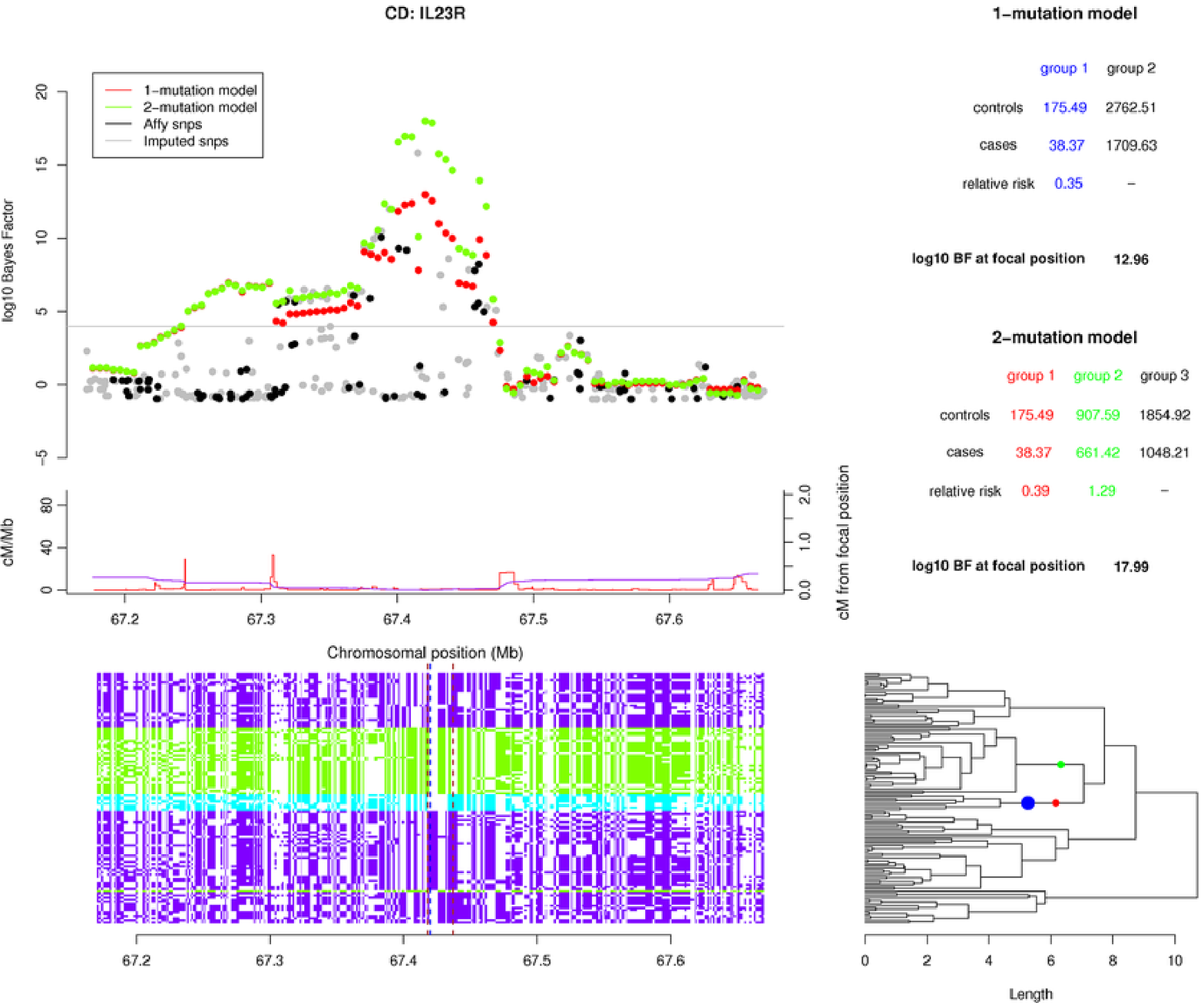}

\caption{The~top left panel of the plot shows the $\log_{10}$ Bayes
factor for the \mbox{1-mutation} model (red) and \mbox{2-mutation} model (green)
within the \textit{IL23R} region of the Crohn's disease analysis. The~recombination
map (red line) and the cumulative recombination map
(purple line) are shown below this. The~bottom left panel shows the 120
CEU HapMap haplotypes across the region. Each row of this panel is a
haplotype and each column is a SNP. The~panel haplotypes are colored to
indicate the three haplotypes that occur at the 2 coding SNPs rs11209026
and rs10889677 (blue${}={}$AC, purple${}={}$GC, green${}={}$GA). The~dashed vertical
blue and brown lines indicate the position of the largest $\log_{10}$
Bayes factor for the \mbox{2-mutation} model (the focal position) and the two
coding SNPs, respectively. The~bottom right panel shows the estimated
genealogical tree at the focal position. The~$x$-axis of the plot was
chosen to provide a clear view of all the branches in the tree. The~branches
associated with the best \mbox{1-mutation} and \mbox{2-mutation} models that
make the largest contributions to the Bayes factors are shown with blue
and red/green dots respectively. The~top right panel shows the tables of
expected allele counts for the \mbox{1-mutation} and \mbox{2-mutation} models together
with a summary of the Bayes factors that occur at the focal position.
The~columns of the tables are color matched to the mutations on the tree
in the bottom right panel.}\label{fig2}
\end{figure*}

There are three regions for the Crohn's disease analysis for which the
posterior probability for two mutations is very close to 1.0. The~first
of these is the \textit{NOD2} region described above. The~second is the
\textit{IL23R} locus on chromosome 1, which is another established
disease locus for Crohn's disease with extensive known allelic
heterogeneity \cite{6}. A plot showing the results of our method in this
region is given in Figure \ref{fig2}. The~$\log_{10}$ Bayes factors, at the
\textit{IL23R} locus, are 12.96 and 17.99 for the \mbox{1-mutation} and
\mbox{2-mutation} models respectively, which compare favorably with the best
Affymetrix SNP (10.07) and the best imputed SNP (15.82). The~difference
between the \mbox{2-mutation} and \mbox{1-mutation} Bayes factors implies a posterior
probability of 1.00 for two mutations versus one mutation, indicating
overwhelming evidence of allelic heterogeneity.

The~original paper \cite{6} identified two SNPs in functional regions of the
\textit{IL23R} gene. The~first SNP\break  (rs11209026) is the nonsynonymous
SNP (c.1142G$>$\break A, p.Arg381Gln) identified as the strongest signal in the
original study. The~second SNP (rs10889677) is in the 3$'$ UTR of the
\textit{IL23R} gene and the only other associated nonintronic SNP found
in the original study. When we look at these two SNPs in the CEU HapMap
panel we identify three distinct haplotypes colored green, purple and
blue in Figure \ref{fig2}. These haplotypes are almost precisely those that are
delineated by the two mutations that make the largest contribution to
the \mbox{2-mutation} Bayes factor. One of the mutations on the tree (colored
red) identifies all the CEU HapMap haplotypes that carry the A allele at
rs11209026 and the second mutation (colored green) identifies all but
one of the haplotypes that carry the A allele at rs10889677. Relative
risk estimates of red and green mutations on the tree, relative to a
lack of either of these mutations, are 0.39 and 1.29 respectively.

Another signal for Crohn's disease is located within an approximately
250~kb region on chromosome 5, flanked by recombination hotspots.
Numerous SNPs within this region have been identified and replicated \cite{20,21} ($p$-values down to 10$^{-12}$ in combined analysis). The~LD
structure delineates this region into five LD blocks and the strongest
associations (single SNP and haplotype) were found in a central 122~kb
block. However, multivariate haplotype analysis conditional on the
effect of the central block showed that the two flanking LD blocks
remain significantly associated \cite{20}, which suggests that multiple
variants in the region may account for the observed effects on Crohn's
disease.

\begin{figure*}[t]

\includegraphics{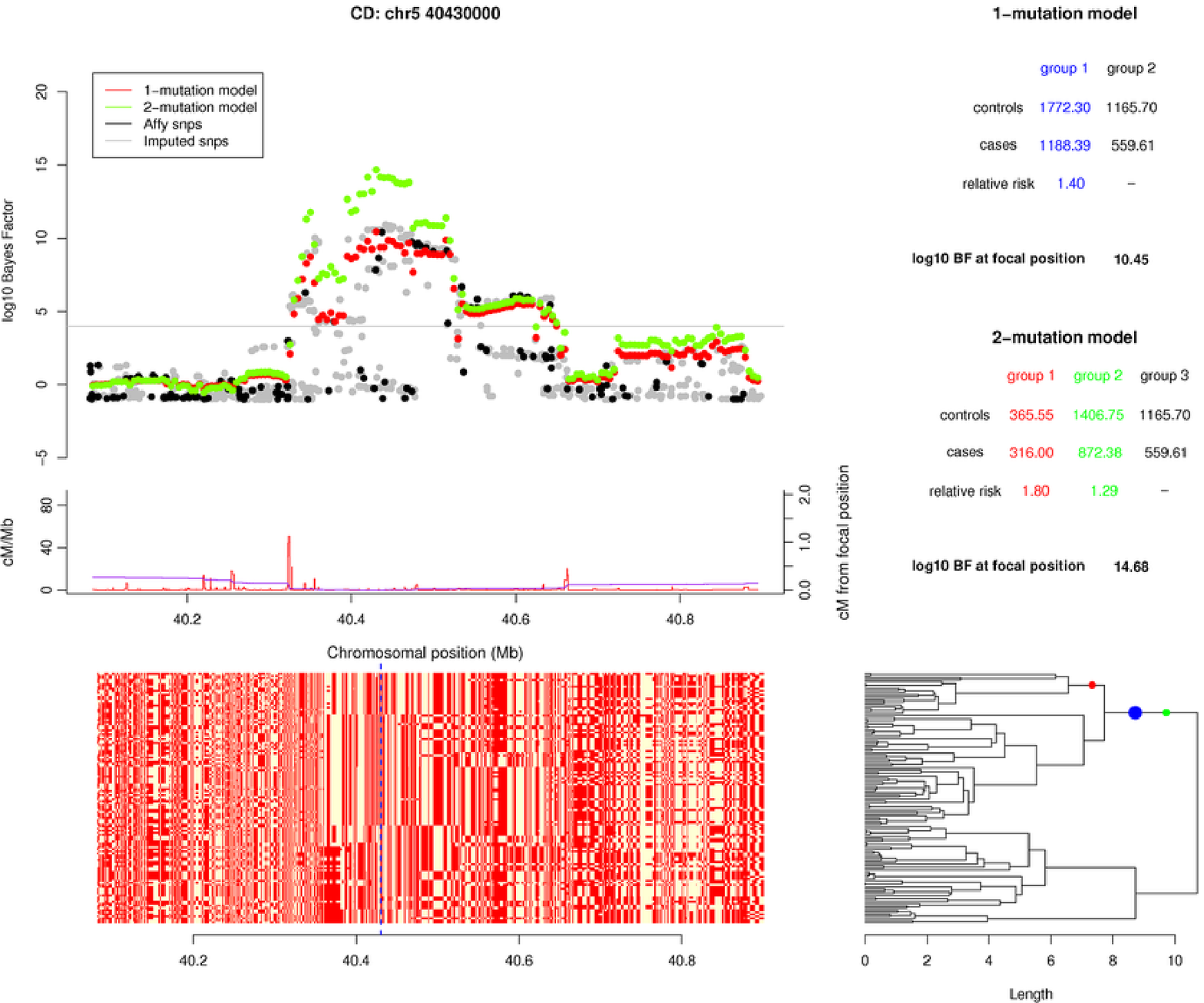}

\caption{The~top left panel of the plot shows the $\log_{10}$ Bayes
factor for the \mbox{1-mutation} model (red) and \mbox{2-mutation} model (green)
within a chromosome 5 region of the Crohn's disease analysis. The~recombination
map (red line) and the cumulative recombination map
(purple line) are shown below this. The~bottom left panel shows the 120
CEU HapMap haplotypes across the region. Each row of this panel is a
haplotype and each column is a SNP. The~panel haplotypes are colored red
and beige to represent the two-allele types at each SNP. The~dashed
vertical blue line indicates the position of the largest $\log_{10}$
Bayes factor for the \mbox{2-mutation} model (the focal position). The~bottom
right panel shows the estimated genealogical tree at the focal position.
The~$x$-axis of the plot was chosen to provide a clear view of all the
branches in the tree. The~branches associated with the best \mbox{1-mutation}
and \mbox{2-mutation} models that make the largest contributions to the Bayes
factors are shown with blue and red/green dots respectively. The~top
right panel shows the tables of expected allele counts for the
\mbox{1-mutation} and \mbox{2-mutation} models together with a summary of the Bayes
factors that occur at the focal position. The~columns of the tables are
color matched to the mutations on the tree in the bottom right panel.}\label{fig3}
\end{figure*}

Single SNP analysis in the WTCCC dataset reveals strong associations at
both Affymetrix and imputed SNPs (maximum $\log_{10}$ Bayes factors 10.41
and 10.92, respectively). Figure \ref{fig3} illustrates the results of our
analysis. The~\mbox{2-mutation} model provides a large boost in signal (maximum
$\log_{10}$ Bayes factor 14.68) and compared to the \mbox{1-mutation} model
(maximum $\log_{10}$ Bayes factor 10.45) strongly support allelic
heterogeneity at this locus (posterior probability of \mbox{2-mutation} model
vs. \mbox{1-mutation} model is 1.0). Further, the two mutations that make the
largest contribution to the \mbox{2-mutation} Bayes factor, appear to delineate
the HapMap haplotypes in three groups with distinct LD pattern
approximately 100~kb either side of the position of the maximum Bayes
factor under the \mbox{2-mutation} model, at 40,430,000 (NCBI Build 35
coordinates), which we call the focal position. Relative risk estimates
of red and green mutations on the tree, relative to a lack of either of
these mutations, are 1.80 and 1.29 respectively.

In addition to the signals identified by the tested typed and imputed
SNPs in the main WTCCC analysis, we find two other signals: one for Type
2 Diabetes (T2D) on chromosome 9 and one for Type 1 Diabetes (T1D) on
chromosome 15.

The~Type 2 Diabetes signal on chromosome 9 resides within a 9~kb region
flanked by recombination hot spots. This locus was identified and
confirmed by three independent T2D genome-wide association
studies \cite{22,23,24}, which reported rs10811661 with the strongest signal of
association. The~$p$-values at this SNP were $7.6\times 10^{-4}$ in the WTCCC
study \cite{24}, $5.4\times 10^{-4}$ in the DGI study \cite{23} and $2.2\times 10^{-3}$ in the
FUSION study \cite{22}. A meta-analysis of the pooled samples from all three
studies \cite{24}, which comprised of 14,586 cases and 17,968 controls,
yielded a $p$-value of $7.8\times 10^{-15}$. A haplotype analysis of this
region also identified a significant signal in this region and the
existence of a high-risk haplotype carrying the T alleles at SNPs
rs10811661 and rs10757283 (see Supplementary Material of ref. \cite{16}).

\begin{figure*}[t]

\includegraphics{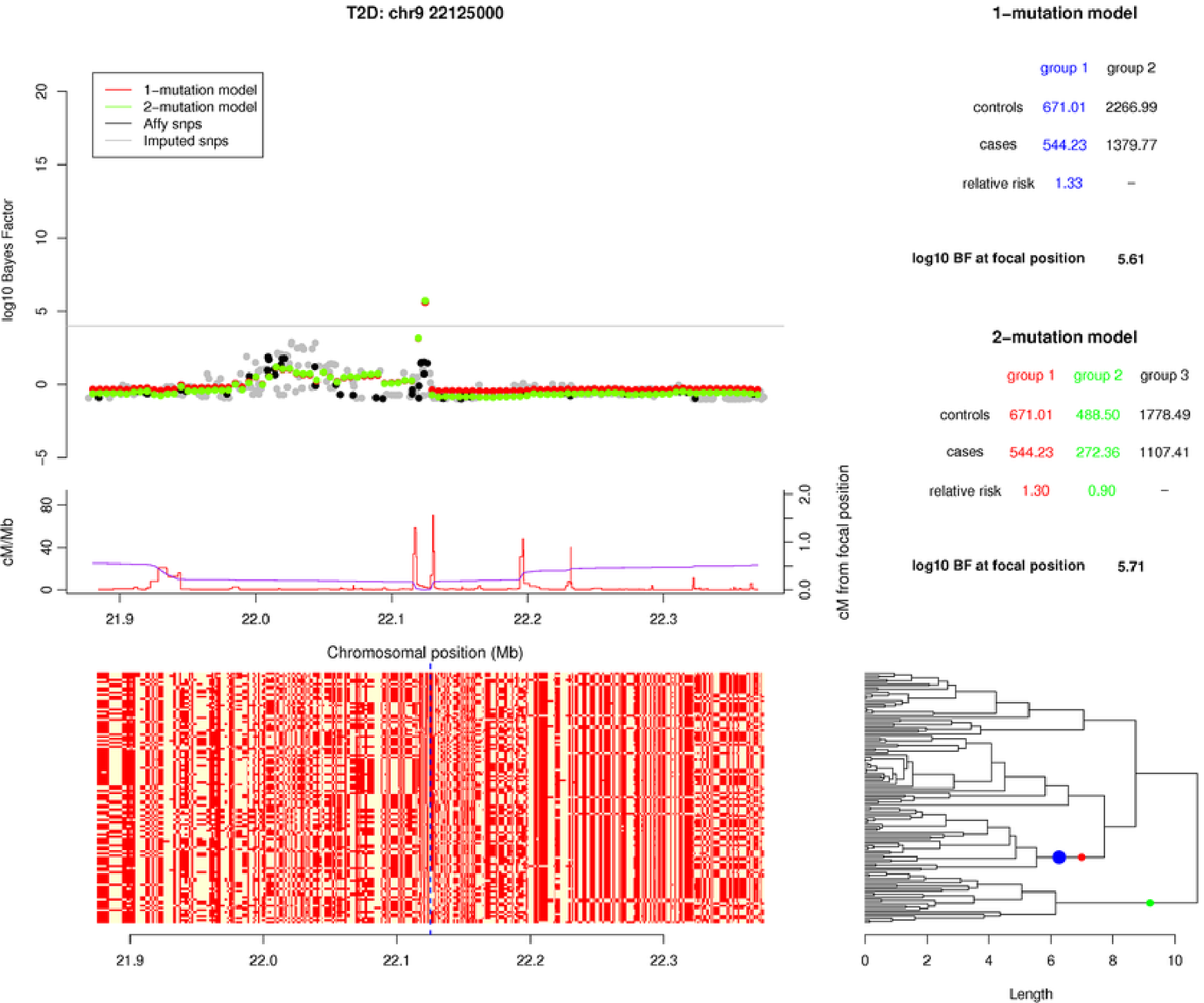}

\caption{The~top left panel of the plot shows the $\log_{10}$ Bayes
factor for the \mbox{1-mutation} model (red) and \mbox{2-mutation} model (green)
within a chromosome 9 region of the Type 2 Diabetes disease analysis.
The~recombination map (red line) and the cumulative recombination map
(purple line) are shown below this. The~bottom left panel shows the 120
CEU HapMap haplotypes across the region. Each row of this panel is a
haplotype and each column is a SNP. The~panel haplotypes are colored red
and beige to represent the two allele types at each SNP. The~dashed
vertical blue line indicates the position of the largest $\log_{10}$
Bayes factor for the \mbox{2-mutation} model (the focal position). The~bottom
right panel shows the estimated genealogical tree at the focal position.
The~$x$-axis of the plot was chosen to provide a clear view of all the
branches in the tree. The~branches associated with the best \mbox{1-mutation}
and \mbox{2-mutation} models that make the largest contributions to the Bayes
factors are shown with blue and red/green dots respectively. The~top
right panel shows the tables of expected allele counts for the
\mbox{1-mutation} and \mbox{2-mutation} models together with a summary of the Bayes
factors that occur at the focal position. The~columns of the tables are
color matched to the mutations on the tree in the bottom right panel.}\label{fig4}
\end{figure*}

Single SNP analyses of the WTCCC data revealed a moderate signal at
rs10811611 of $\log_{10}$ Bayes factor 1.53, which is the strongest
within the 9~kb region flanking the recombination hotspots (stronger
signals are located approximately 100~kb away but are likely to be
related to another signal associated with rs564398). Figure \ref{fig4} summarizes
our results in this region. The~maximum $\log_{10}$ Bayes factors peak at
5.61 and 5.71 for the \mbox{1-mutation} and \mbox{2-mutation} models. These signals
represent a significant boost in power to detect this locus. The~\mbox{2-mutation} model provides a better fit than the \mbox{1-mutation} model
suggesting evidence of allelic heterogeneity in the region. Relative
risk estimates of red and green mutations on the tree, relative to a
lack of either of these mutations, are 1.30 and 0.90 respectively.

One of the mutations on the tree (colored red) exactly identifies all
but one of the HapMap haplotypes that contain the high risk TT haplotype
at SNPs rs10811661 and rs10757283. The~other mutation on the tree
(colored green) identifies a protective CT haplotype at the SNPs
rs10811661 and rs10757283 that was not mentioned in the original
analysis.

\begin{figure*}[t]

\includegraphics{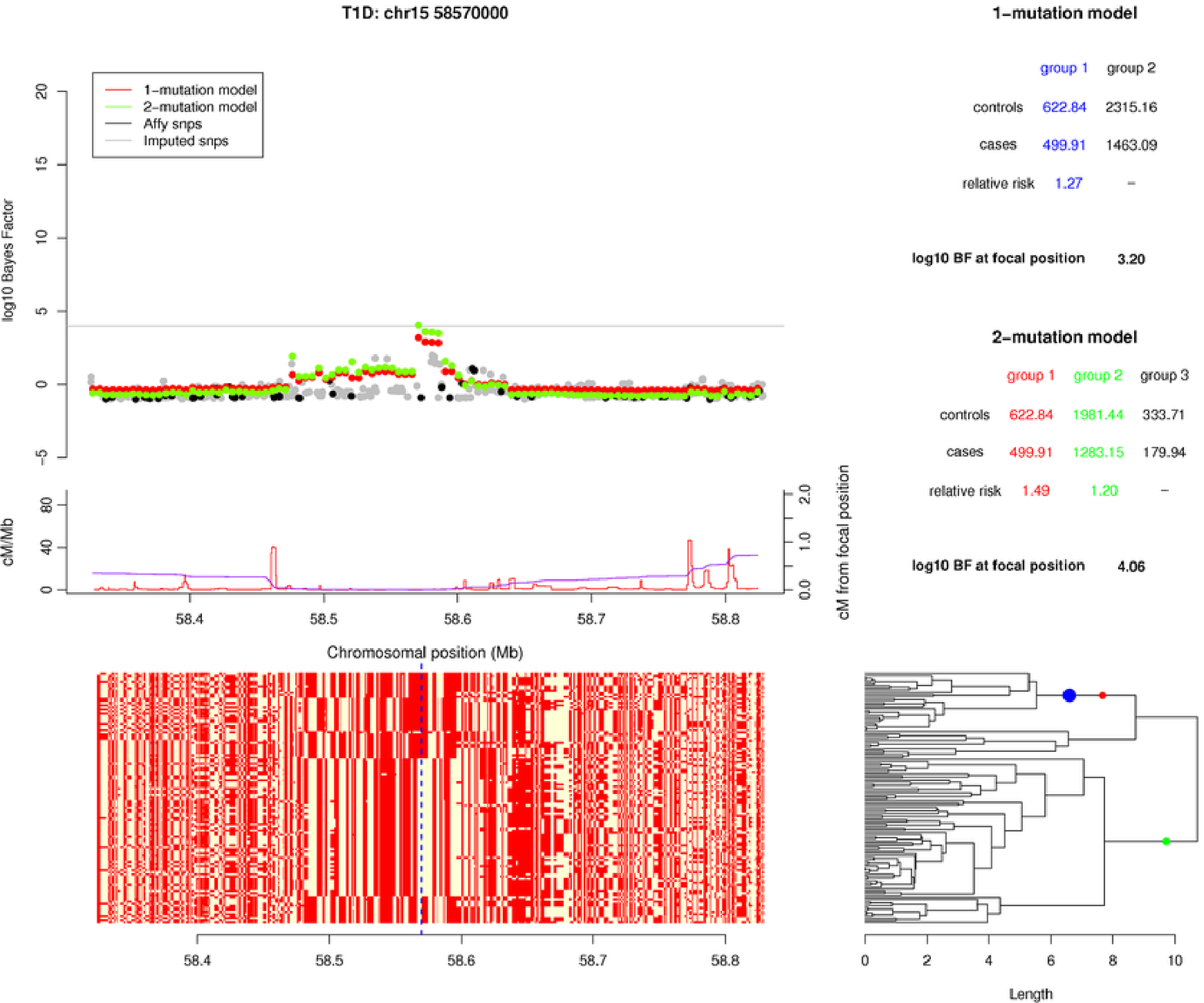}

\caption{The~top left panel of the plot shows the $\log_{10}$ Bayes
factor for the \mbox{1-mutation} model (red) and \mbox{2-mutation} model (green)
within a chromosome 15 region of the Type 1 Diabetes disease analysis.
The~recombination map (red line) and the cumulative recombination map
(purple line) are shown below this. The~bottom left panel shows the 120
CEU HapMap haplotypes across the region. Each row of this panel is a
haplotype and each column is a SNP. The~panel haplotypes are colored red
and beige to represent the two allele types at each SNP. The~dashed
vertical blue line indicates the position of the largest $\log_{10}$
Bayes factor for the \mbox{2-mutation} model (the focal position). The~bottom
right panel shows the estimated genealogical tree at the focal position.
The~$x$-axis of the plot was chosen to provide a clear view of all the
branches in the tree. The~branches associated with the best \mbox{1-mutation}
and \mbox{2-mutation} models that make the largest contributions to the Bayes
factors are shown with blue and red/green dots respectively. The~top
right panel shows the tables of expected allele counts for the
\mbox{1-mutation} and \mbox{2-mutation} models together with a summary of the Bayes
factors that occur at the focal position. The~columns of the tables are
color matched to the mutations on the tree in the bottom right panel.}\label{fig5}
\end{figure*}

A possible novel signal is located at chromosome 15q22.2 for T1D (Figure
\ref{fig5}), where no previous associations have been identified. Single SNP
tests only detected a very weak signal in this region ($\log_{10}$ Bayes
factor peak at 1.08 and 1.98 at Affymetrix and imputed SNPs
respectively). The~maximum $\log_{10}$ Bayes factor from the \mbox{2-mutation}
model (4.06) is stronger than the \mbox{1-mutation} model (3.20), which
provides some suggestion that multiple causal variants are involved. The~focal position of our signal is located in the \textit{RORA} gene, which
encodes \textit{ROR}, an evolutionarily related transcription factor and
belongs to the steroid hormone receptor super family. \textit{RORA} has
been linked to immunomodulatory activities \cite{25}, which might make
\textit{RORA} a candidate gene for autoimmune diseases such as T1D.

\begin{figure*}[t]

\includegraphics{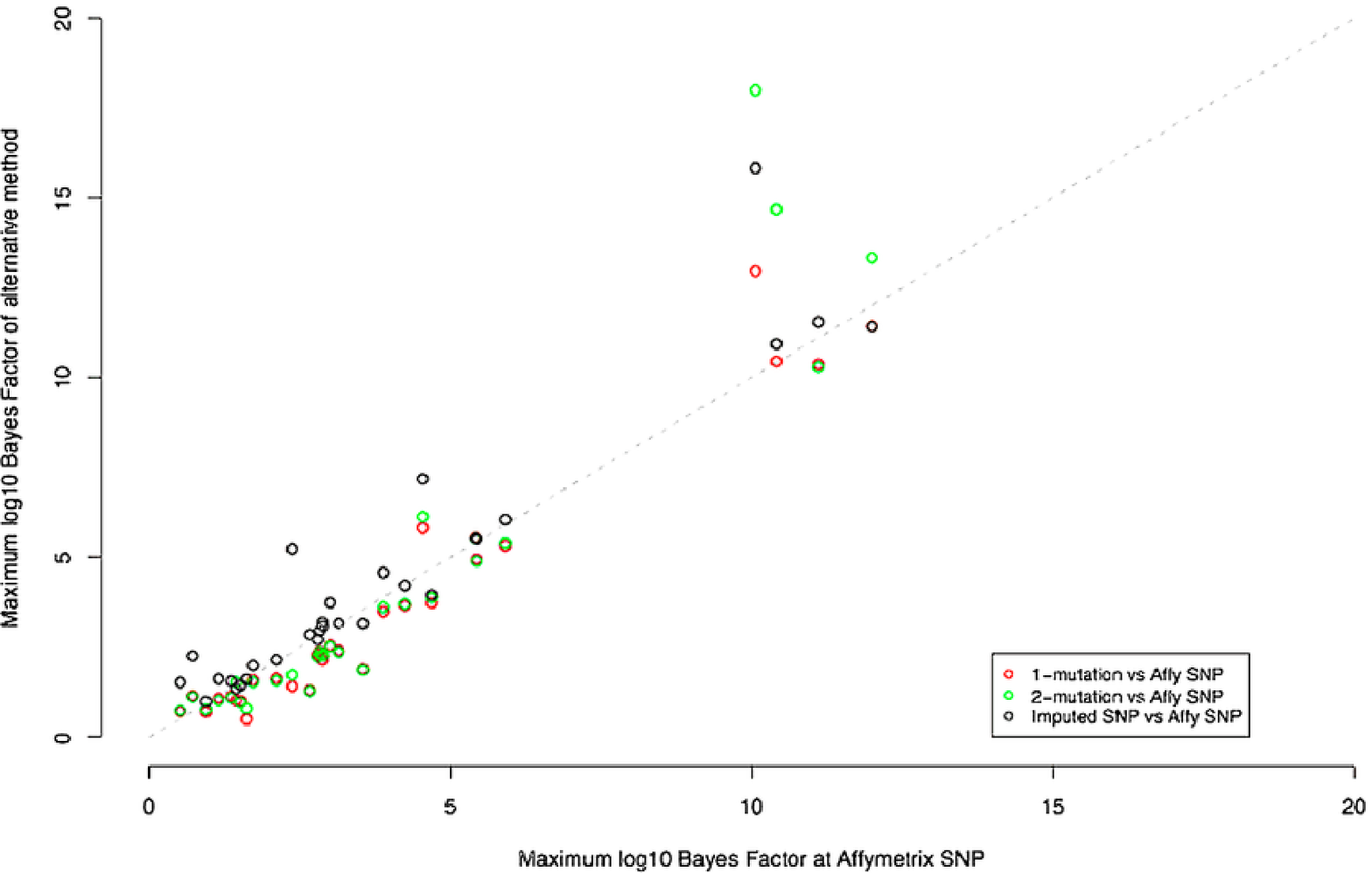}

\caption{Comparison of the performance of four different methods in the
WTCCC data at the 30 established associated loci for Crohn's disease.
The~plot shows the maximum $\log_{10}$ Bayes factor for imputed SNPs
(black) and the \mbox{1-mutation} (red) and \mbox{2-mutation} (green) versions of our
new method (on the $y$-axis), plotted against the maximum $\log_{10}$ Bayes
factor at Affymetrix SNPs (on the $x$-axis), in each region.}\label{fig6}
\end{figure*}

\begin{figure*}[t]

\includegraphics{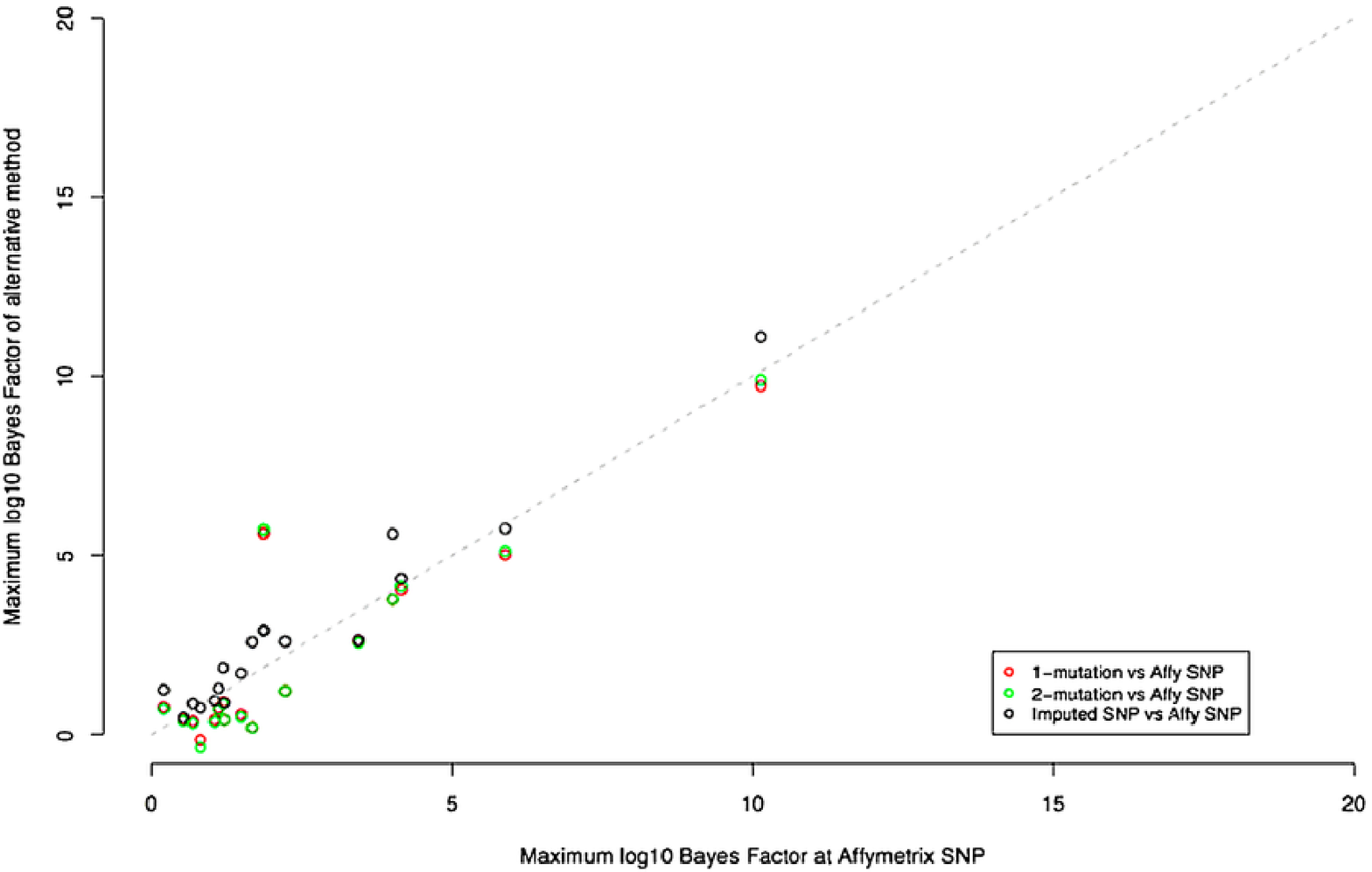}

\caption{Comparison of the performance of four different methods in the
WTCCC data at the 18 established associated loci for Type~2 Diabetes.
The~plot shows the maximum $\log_{10}$ Bayes factor for imputed SNPs
(black) and the \mbox{1-mutation} (red) and \mbox{2-mutation} (green) versions of our
new method (on the $y$-axis), plotted against the maximum $\log_{10}$ Bayes
factor at Affymetrix SNPs (on the $x$-axis), in each region.}\label{fig7}
\end{figure*}

We also looked at the WTCCC data in detail at the relatively large
number of established disease genes for Crohn's disease \cite{21} (30 loci)
and Type 2 Diabetes \cite{26} (18 loci). Not all of these loci were found to
be highly significant in the WTCCC study. We compared tests at (i) SNPs
on the Affymetrix 500k chip, (ii) imputed SNPs, and the (iii) \mbox{1-mutation}
and (iv) \mbox{2-mutation} versions of our new method. The~results for the
Crohn's Disease and Type 2 Diabetes regions are shown in Figures \ref{fig6} and \ref{fig7}
respectively. The~results show that no one method uniformly produces the
largest signal across all of the regions. Out of all the 48 regions
together, the four methods produced the largest signal in 12, 30, 1 and
5 regions respectively. These results show that in 13\% of the regions
of known association the methods described in this paper lead to an
increase in signal over and above that of testing directly typed and
imputed SNPs (although in some cases the increase in signal is small).
The~results also reinforce our previous findings \cite{8} that imputation can
provide a nontrivial boost in power over testing only those SNPs that
have been genotyped directly.

\subsection{Power to Detect Allelic Heterogeneity}

The~real examples shown above illustrate that when allelic heterogeneity
exists in a region of association our method can accurately characterize
the underlying risk variants and lead to a boost in signal. We have also
used simulation to assess the power of our method to distinguish the
signal of allelic heterogeneity when it exists. To do this we extended
our program HAPGEN \cite{27} to simulate SNP genotype datasets, in 2000 cases
and 2000 controls, under a model of allelic heterogeneity with two
linked causal SNPs. HAPGEN conditions on a reference panel of haplotypes
and their local recombination rates to create genotype datasets that
naturally inherit the patterns of LD found in the reference panel.
Datasets were generated by using the 120 CEU\break parental HapMap haplotypes
in five ENCODE regions (ENr123, ENr213, ENr232, ENr321 and\break ENm013) as
reference panels; each dataset is approximately 500~kb in length with a
SNP density of approximately two SNPs per kb. For the set of haplotypes
required by step~1 of our method we produced pseudo-HapMap panels by
thinning the ENCODE data to match the SNP density and MAF distribution
of the phase II HapMap data, with the added restriction that this panel
contain the SNPs on the Affymetrix 500k~chip. Each dataset was simulated
at all SNPs in the ENCODE regions but only genotype data at the SNPs on
the Affymetrix 500k chip were presented to the method in step 2.

To simulate instances of allelic heterogeneity we selected pairs of SNPs
within 15~kb of each other, and satisfying the condition of either Model
A or Model B (described below), as the causal SNPs for a dataset. We
exhaustively searched for all suitable pairs in the five ENCODE marker
sets and for each pair generated a single dataset, comprised of 2000
case and 2000 control individuals. The~minor allele was set to be the
deleterious allele at both SNPs and phenotypes were simulated according
to the marginal relative risks given to each disease allele.

\begin{table*}
\caption{Results of simulations of allelic heterogeneity at two linked
causal SNPs using Model A: one rare with risk allele frequency less than
2\% and one common with risk allele frequency between 5\% and 20\%.
Relative risks at the two simulated loci are shown in the first two
rows. The~maximum \mbox{1-mutation} and \mbox{2-mutation} $\log_{10}$ Bayes factors are
denoted by $S_{1}$ and $S_{2}$ respectively. The~third row shows the
proportion of~simulated datasets where $S_{2}$ was greater than $S_{1}$.
The~fourth row shows the proportion of simulated datasets that had
$S_{2} > 3$. The fifth~row~shows the proportion of simulated datasets
where $S_{2}$ was greater than $S_{1}$ conditional upon $S_{2} > 3$. The~final
row shows the~expected~difference~between~$S_{2}$ and $S_{1}$
conditional upon $S_{2} > 3$}
\label{tab2}
\begin{tabular*}{\textwidth}{@{\extracolsep{\fill}}lccccc@{}}
\hline
RR$_{\mathrm{A}}$ (rare causal SNP) & 1.0\phantom{0} & 1.0\phantom{0} & 1.5\phantom{0} & 2.0\phantom{0} & 2.5\phantom{0}\\
RR$_{\mathrm{B}}$ (common causal SNP) & 1.0\phantom{0} & 1.3\phantom{0} & 1.3\phantom{0} & 1.3\phantom{0} & 1.3\phantom{0}\\
$\operatorname{Pr}(S_{2} > S_{1})$ & 0.07 & 0.33 & 0.45 & 0.61 & 0.70\\
$\operatorname{Pr}(S_{2} > 3)$ & 0.00 & 0.13 & 0.18 & 0.37 & 0.57\\
$\operatorname{Pr}(S_{2} > S_{1} | S_{2} > 3)$ & --- & 0.52 & 0.69 & 0.81 & 0.83\\
$\operatorname{Mean}(S_{2} - S_{1} | S_{2} > 3)$ & --- & 0.07 & 0.20 & 0.59 & 0.73\\
\hline
\end{tabular*}
\end{table*}

\begin{table*}[b]
\caption{Results of simulations of allelic heterogeneity at two linked
causal SNPs using Model B: both causal SNPs with a risk allele frequency
between 5\% and 20\%. Relative risks at the 2 simulated loci are shown
in the first two rows. The~maximum \mbox{1-mutation} and \mbox{2-mutation}
$\log_{10}$~Bayes factors are denoted by $S_{1}$ and $S_{2}$ respectively. The~third
row shows the proportion of simulated datasets where $S_{2}$ was
greater~than~$S_{1}$. The~fourth row shows the proportion of simulated datasets
that had $S_{2} > 3$. The~fifth row shows the proportion of~simulated~datasets
where $S_{2}$ was greater than $S_{1}$ conditional upon $S_{2}
> 3$. The~final row shows the expected difference between~$S_{2}$~and~$S_{1}$ conditional~upon~$S_{2} > 3$}
\label{tab3}
\begin{tabular*}{\textwidth}{@{\extracolsep{\fill}}lccccc@{}}
\hline
RR$_{\mathrm{A}}$ & 1.0\phantom{0} & 1.0\phantom{0} & 1.1\phantom{0} & 1.3\phantom{0} & 1.5\phantom{0}\\
RR$_{\mathrm{B}}$ & 1.0\phantom{0} & 1.3\phantom{0} & 1.3\phantom{0} & 1.3\phantom{0} & 1.3\phantom{0}\\
$\operatorname{Pr}(S_{2} > S_{1})$ & 0.05 & 0.32 & 0.38 & 0.55 & 0.67\\
$\operatorname{Pr}(S_{2} > 3)$ & 0.00 & 0.21 & 0.33 & 0.56 & 0.81\\
$\operatorname{Pr}(S_{2} > S_{1} | S_{2} > 3)$ & --- & 0.44 & 0.47 & 0.56 & 0.69\\
$\operatorname{Mean}(S_{2} - S_{1} | S_{2} > 3)$ & --- & 0.04 & 0.04 & 0.17 & 0.41\\
\hline
\end{tabular*}
\end{table*}

For each simulated dataset we compared the maximum \mbox{1-mutation} and
\mbox{2-mutation} Bayes factors. Tables~\ref{tab2} and~\ref{tab3} show the results of this
comparison for two disease models that we simulated: Model A, one rare
causal SNP with risk allele frequency less than 2\% and one common
causal SNP with risk allele frequency between 5\% and 20\%, and Model B,
two causal SNPs with a risk allele frequency between 5\% and 20\%.

The~tables show the proportion of times that the \mbox{2-mutation} signal is
larger than that for the \mbox{1-mutation} model. We also show results for just
those simulated datasets where there is an appreciable signal of
association ($\log_{10}$ Bayes factor $>$ 3). In general, these results
show that our method has good power to detect allelic heterogeneity when
the effect sizes at the susceptibility loci are similar to those found
in our analysis of the WTCCC data. For example, when the relative risks
are 2.5 and 1.3 at the rare and common SNPs for Model~A our method has
70\% power to detect a larger signal for the \mbox{2-mutation} model. If we
consider only those simulations in which the signal is appreciably large
($\log_{10}$ Bayes factor $>$ 3) then this power rises to 83\%. Similarly
for Model B, when the relative risks are 1.5 and 1.3 at the
susceptibility SNPs our method has 67\% power to detect a larger signal
for the \mbox{2-mutation} model and this power rises to 69\% when conditioning
only on large signals. As effect sizes become smaller there is less
power to detect an effect and it also becomes more difficult to
distinguish between one and two mutations. When there is no effect at
either locus, that is, under the ``null hypothesis'' of no association, we
obtain a false positive rate of close to zero when conditioning upon
appreciable signals.

\subsection{Estimating Effect Sizes in Associated Regions}

In associated regions it is standard practice to report the effect size
of the risk allele at the associated SNP and it is usual that this takes
the form of the estimated Relative Risk (RR) or Odds Ratio (OR) of the
allele together with a 95\% confidence interval. Such estimates are
useful for approximating the magnitude and precision of the association
in the study population, quantifying the amount of heritability
explained by the locus and predicting individual disease risk. As we
have seen above, testing for association at typed and imputed SNPs can
be successful in detecting associated regions but this is not always the
case and our method is sometimes able to detect a larger signal,
effectively by more accurately characterizing the true causal variant.
It follows that in these cases our method may also be able to accurately
estimate the effect size of the true causal variant. To investigate this
idea we carried out a simulation study using the ENCODE region ENm013
from the CEU HapMap haplotypes and the thinned pseudo-HapMap panel that
we created for our simulation study in the previous section. We searched
for all SNPs in the ENCODE region that had an $R^{2}$ with any SNP in
the pseudo-HapMap panel of at most 0.2 and used these SNPs as the causal
SNPs in our simulations. These SNPs will be not be in high LD with any
of the SNPs on the Affymetrix 500k chip and are unlikely to be imputed
well. For each causal SNP we then simulated a case-control study in the
region using HAPGEN. Each causal SNP was used four times with simulated
relative risks of 1.25, 1.5, 2.0 and~2.5. Only genotype data at the SNPs
on the Affymetrix 500k chip were simulated. We then analyzed the data in
two different ways to obtain an approximate posterior distribution on
the effect size.

Firstly, we considered the estimated effect size at the most associated
Affymetrix SNP in the region. We used a logistic regression model and
fitted an additive model on the log odds scale implemented by SNPTEST to
calculate the mode of the posterior distribution of additive effect
parameter, $\hat{\beta}$. The~OR estimate is subsequently calculated
as $e^{\hat{\beta}}$. The~prior on the effect size was that used in the
WTCCC study \cite{5}.

We then obtained an analogous estimate and standard errors from our
method GENECLUSTER in the following way. We first identified the
locus $X_{m}$, where the maximal \mbox{1-mutation} Bayes factor occurred. For
each branch, $b$, on the genealogical tree constructed at this
position we placed a mutation on the branch and calculated the posterior
probability that the $i$th individual carried 0, 1 or 2 copies of
the mutation, $P( G_{i}^{mb}| H^{mb})$. We then took these genotype
distributions at all individuals and used SNPTEST to carry out a test of
association at the SNP implied by the mutation on the branch using the
same additive logistic regression model as above. This resulted in a
posterior estimate of $\beta_{b}$ and its standard
error $\sigma_{b}^{2}$. The~posterior distribution can be calculated by
summing over the branches of the tree, that is,
\begin{eqnarray*}
 P( \beta |\mathit{Data} ) &=& \sum_{b} P( \beta,b|\mathit{Data} )\\
  &= &\sum_{b} P( \beta |b,\mathit{Data} )P(b|\mathit{Data} ) \\
  & \propto& \sum_{b} P( \beta |b,\mathit{Data} )P(\mathit{Data}|b )P( b ) \\
  & \propto& \sum_{b} P( \beta|b,\mathit{Data} )\mathit{BF}_{b}P( b ) ,
\end{eqnarray*}
where $\mathit{BF}_{b}$ is the Bayes factor associated with branch $b$ and $P( b
)$ is the prior probability on branch $b$ carrying a causal mutation. If
we assume that the posterior distribution of the additive genetic effect
parameter conditional on a given branch, $P( \beta |b,\mathit{Data} )$, can be
approximated using a Normal distribution $N( \hat{\beta}
_{b},\hat{\sigma} _{b}^{2} )$ then the overall estimate will be a
mixture of Normal distributions with each branch weighted by its
associated Bayes factor and its prior. From this model we can obtain a
new estimate of the effect size as
\[
\hat{\beta} ^{*} = \frac{1}{K}\sum_{b} \hat{\beta} _{b}\mathit{BF}_{b}P( b ),
\]
where $K = \sum_{b} \mathit{BF}_{b}P( b )$. The~OR estimate is subsequently
calculated as $e^{\hat{\beta} ^{*}}$.

\begin{figure*}

\includegraphics{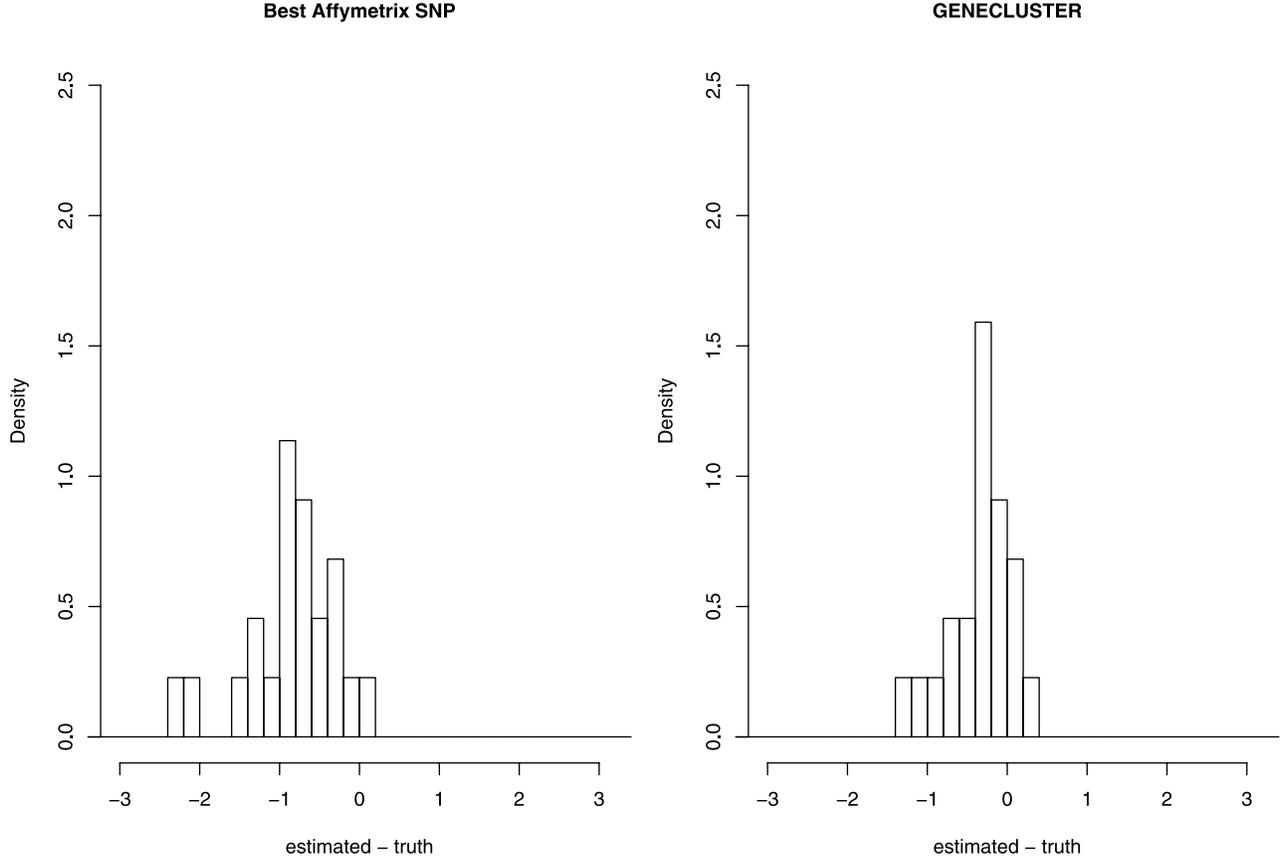}

\caption{Comparison of methods for estimating the effect size. Both
plots show the distribution of the difference between the estimated and
true odds ratio. The~left hand plot shows the results when using the
best chip SNP to estimate odds ratio. The~right-hand plot shows the
results when using GENECLUSTER.}\label{fig8}
\end{figure*}

We compared these two estimates of the effect size to the true estimate
of the effect size, which we calculated by fitting the same logistic
regression model to the simulated data at the true causal SNP. Figure \ref{fig8}
shows the distribution of the difference between the estimated effect
size minus the true effect size for both methods. In constructing this
plot we only considered simulations that showed a maximal $\log_{10}$
Bayes factor for the \mbox{1-mutation} GENECLUSTER model above 4. The~plot
shows that GENECLUSTER outperforms the use of the best Affymetrix SNP
when estimating the effect size. The~mean square error for the OR
estimate is 1.037 for the best Affymetrix SNP estimate and 0.524 for the
GENECLUSTER method.

\section{Discussion}

The~standard paradigm for the analysis of genome-wide association
studies involves testing both typed and imputed SNPs and then attempting
to replicate interesting signals in new datasets. In this paper, we have
proposed a complementary method that attempts to extract further signals
of association first by explicitly considering as-yet-unknown SNPs in
the region, and second by modeling and estimating allelic
heterogeneity at a locus. Allelic heterogeneity has been predicted to
play a significant role in the genetic etiology of complex diseases \cite{28}
and clear examples in real human data already exist (Figures \ref{fig1}--\ref{fig3}). Our
method works by locally approximating the genealogy of the haplotypes in
the sampled individuals and then averaging over the different branches
of the genealogy as potential sites of casual mutations using a Bayesian
approach.

A key feature of our approach is the use of a genealogical tree to
represent the relationship between the haplotypes of the sample and to
effectively constrain the space of possible causal variants considered.
The~genealogical tree, built in step 1 using fine-scale haplotype data
at each position, greatly aids interpretation of the signal. As
illustrated in Figures \ref{fig1}--\ref{fig5}, we are able to accurately estimate the best
single branch, and pair of branches on the tree, that make the largest
contribution to the signal of association. Since the tree is built using
only the set of HapMap haplotypes we are able to graphically link the
tree to the haplotypes themselves, which acts to highlight the
haplotypic backgrounds that harbor the estimated causal mutations. Our
analyses in this paper are based on a set of genealogical trees built at
5~kb intervals and testing for association at those locations. The~5~kb
interval size was chosen based on the SNP densities of our data in the
HapMap reference panel, which is approximately one SNP per kb, and in
the study sample, which is approximately one SNP per 6~kb. Using an
excessively small interval size compared to the SNP densities in the
reference panel and study sample will likely yield highly a set of
 correlated trees (in step 1), clusterings of the study sample
genotypes (in step 2) and hence signals for association (in step 3).
However, some brief analyses indicate that 5~kb could be a conservative
estimate and a higher density, for example, one tree per 1~kb, can
increase power and lead to a finer resolution for the location of the
putative disease locus (results not shown). There is little disadvantage
in testing at more locations, apart from the linear increase in the
computation burden, and since modern association study data are typed at
an increasingly dense set of SNPs, we recommend implementing GENECLUSTER
using as dense a set of trees as computation resources allow.

For the analysis in this paper our method was applied using one
estimated genealogy at each position on a grid of positions across the
genome. By using a single tree at each position it is straightforward to
visualize which branches, or combination of branches, drive the signal
of association. The~disadvantage is that the uncertainty in the
genealogy is ignored. Our method is currently being extended to allow
multiple trees at each position in order to capture the uncertainty in
the genealogy but it is not clear whether this will lead to a
significant boost in performance and we have left this for future work.
However, we feel encouraged by the performance of our current method in
its ability to accurately identify known allelic heterogeneity and to
boost signals of association in real data and simulated data (see
Supplementary Material).

The~second step of our method involves locally clustering the genotype
data to the tips of the estimated genealogy. A key feature of our model
is that we are not constrained to choose a ``window'' of SNPs, as
required by many haplotype clustering methods \cite{29,30,31}, and instead we are
able to use hundreds of flanking SNPs around the focal locus to both
build the genealogical trees and cluster the genotype data to the tips
of the tree. Our method also naturally handles missing data and takes
haplotype uncertainty into account and thus avoids relying on a point
estimate of haplotypes \cite{10} as this has been shown to produce nonoptimal
results \cite{32}. We encourage the use of the most accurate recombination map
possible but experience using similar HMM models for imputation suggests
that the models are reasonably robust to varying the recombination
rates. The~mutation rates in our models are fixed and constant across
SNPs. It could be argued that estimating them in a SNP-by-SNP fashion
might help down weight the influence of SNPs with high genotyping error
rates, but this would add considerably to the computational expense of
the method and since genotyping error rates are low we do not think this
would make a noticeable improvement to our method.

The~third step in our method involves placing one or more mutations on
the estimated genealogy and evaluating the evidence of association
between those mutations and the phenotype data of those genotypes
clustered to the tips of the tree. We set our prior probability of a
mutation occurring on a given branch to be proportional to the expected
branch length, which is based on the assumption that mutations are more
likely to occur on longer branches and every mutation has an equal prior
probability of being causal. This means that our prior probability on
rare mutations being causal will be small since they tend to occur on
shorter branches. We can also adjust our prior to favor mutations that
occur on the shorter branches to boost our power to detect rare
variants. However, our ability to detect rare causal variants will also
be limited by the characterization of rare variants in the HapMap
reference panel (as discussed below).

Thus far we have only considered placing at most two mutations on a
single genealogy but our approach can be easily extended to placing
further mutations. However, there is a considerable increase in
computation burden with placing more mutations due to the increase in
the set of possible combinations of branches carrying a mutation. For
example, the complexities of the 3-mutation and the 4-mutation models
increase by approximately 79 and 4600 times, respectively, compared to
the \mbox{2-mutation} model. It is therefore feasible to implement the
3-mutation model for analyses of small regions, for example for
fine-mapping, but not genome-wide. A possible compromise would be to
employ a Markov chain Monte Carlo approach to integrate over the space
of branches carrying a mutation.

A key approximation that we make, which is worthy of some discussion, is
to construct the genealogical tree using only the reference sample of
HapMap haplotypes and then probabilistically cluster the study
individuals under the tips of the tree at each locus. In doing this we
effectively construct a genealogical tree for the whole sample. In
contrast, the MARGARITA method \cite{32} attempts to construct the full
genealogy of the sample but only in the study individuals and using a
rather heuristic method for implicitly phasing the genotype data. In
doing so this method ignores the information available from the HapMap
haplotypes in a given region.

The~advantages of using the HapMap data are that the haplotypes are
accurately phased and consist of a higher SNP density than commercially
available genotyping chips. Both of these properties aid the
reconstruction of the genealogical tree. In addition there are
computational advantages in being able to produce a set of genealogies
across the genome just once and then storing the trees for all future
use. A~limiting feature of the HapMap haplotypes is the relatively small
size of the sample, that is, there are only 120 CEU haplotypes. In many
regions of the genome the HapMap haplotypes will provide a good
representation of the common set of haplotypes likely to be found in the
population but there will clearly be regions where this is not true.
Also, the HapMap will not provide a comprehensive characterization of
the rare haplotype structure present in the population. It is clear that
there are instances in which our method of building genealogies will not
be perfect but the application to seven genome-wide scans of the WTCCC
has clearly shown that the method is able to detect and accurately
characterize real associations where they occur. This is likely due to
the fact that it is common variants that show association at these loci
and our method is able to accurately characterize the relevant common
haplotype structure.

We have carried out a small amount of comparison between GENECLUSTER and
MARGARITA on selected loci. At the chromosome 9 locus for Type 2
Diabetes discussed above and shown in Figure \ref{fig4} we found that MARGARITA
was not able to uncover a significant association (permutation $p$-value
0.2498), whereas significant signals at other loci examined in this
paper were found. A more comprehensive comparison of these methods is
complicated by the fact that MARGARITA produces results in terms of
$p$-values whereas GENECLUSTER's inference is Bayesian.

At loci where there is an especially large genetic effect the true
underlying genealogy of the study sample may differ quite a lot to that
of the genealogy of the sample of HapMap haplotypes. In this scenario
the case haplotypes will be strongly clustered under the branch of the
tree that contains the disease susceptibility mutation whereas control
haplotypes will tend to be biased away from clustering under this
branch. Thus in this case using the HapMap haplotypes to build a
genealogy may not be optimal. As the effect size gets smaller, however,
this bias is reduced and we do not see this as a serious concern for
analysis of genome-wide association studies where effect sizes are
typically small.

A more complete method would involve the construction of the
genealogical tree at each position using all the data. This is
complicated by the fact that the study individuals consist of unphased
genotype data, whereas the HapMap haplotypes are phased, and consist of
missing data at many SNPs that are in HapMap but not on the genotyping
chip. One can envisage an iterative scheme in which phasing and
imputation of missing alleles in the study individuals and building of
genealogical trees are carried out, but this would likely be
computationally prohibitive, unless other simplifying assumptions are
made. Strictly speaking it would also be necessary to build the
genealogical tree and fit a disease model at the same time and this
would add a further layer of complexity.

We expect that the performance of our method will show a similar pattern
of variation to that of imputation when applied in other populations
 \cite{33} since the underlying models are quite similar. Applying the method
to admixed individuals or to studies involving individuals from
different populations is not something we have considered here and we
would encourage caution in directly applying the method in such
situations. This may be an interesting avenue for future research.

We see two possible ways in which our method could be used. First, and
foremost, we see it as a complementary method to testing typed and
imputed SNPs across the genome. The~method is designed to pick up
signals that have a more complex structure than ones single SNP models
can accommodate. Our results on the WTCCC datasets above show that the
method is able to boost signal in regions where this occurs. For
example, in the 48 established regions of association for Crohn's
disease and Type 2 Diabetes our new methods produced the largest signal
in 13\% of the regions. A~distinction between our approach and the
SNP-based approaches is that we jointly assess the data at all SNPs
compatible with the genealogy for evidence of association. Therefore, at
each location, GENECLUSTER assesses the evidence for association at
\textit{any} SNP, whereas SNP-based approaches perform a single test at
\textit{each} SNP for association. This means that in regions with a SNP
(typed or imputed) that is either causal, or in strong LD with a causal
SNP, GENECLUSTER is likely to produce a lower Bayes factor than a direct
test at that SNP, and we expect that this is the case for most of the
regions in our comparison since they were identified using SNP-based
approaches. However, our results also show that there is an appreciable
number of regions in the genome where GENECLUSTER outperforms SNP-based
approaches, namely regions with a causal variant that is not well tagged
by the data, or with multiple causal variants. The~Supplementary
Material details a further simulation study that we have carried out to
show that our method is well powered to detect signals of association
compared to simpler tag-based approaches.

Our use of a Bayesian framework allows the results of a GENECLUSTER
analysis to be naturally combined together with the analysis of imputed
and typed SNPs. The~Bayes factors from each approach can be combined
together into one set across the genome, and interesting signals can be
identified by applying a Bayes factor threshold that is determined by
the prior probability of an association. This prior probability
represents the proportion of the genome that we expect to be associated
with the disease, which remains fixed and independent of the number of
tests carried out. This means that our method can be naturally and
easily accommodated into the analysis without recourse to Frequentist
multiple testing procedures. In our analysis, we have assumed
1$/$10,000th of the genome is truly associated  \cite{5}. Determining a
threshold for the Bayes factor involves the use of decision theory and
the specification of a loss function. When focusing on identifying a set
of SNPs for follow-up replication, we might penalize a false
nondiscovery more than a false \mbox{discovery}. When making a final decision
on a SNP after replication data has been collected, we might penalize a
false discovery more than a false nondiscovery. To illustrate our
method and compare methods we have used a 0/1 loss function that gives
equal weight to false discoveries and false nondiscovery and represents
the middle ground between these two scenarios. This results in a common
threshold of 4 for the $\log_{10}$ Bayes factors at typed SNPs, imputed
SNPs and GENECLUSTER. We do not expect the comparison between methods to
be influenced by this choice.

Our method could be used in a more focused fashion, in fine-mapping
experiments, to investigate the form of the association in regions
already identified by single SNP methods and to produce better estimates
of effect sizes. For example, if the application of the \mbox{1-mutation}
version of the method leads to a clear boost in signal over typed and
imputed SNPs then this may indicate the presence of an undiscovered
causal SNP. Further application of the \mbox{2-mutation} version may
subsequently indicate a much stronger signal implying allelic
heterogeneity within the region, such as at the \textit{NOD2},
\textit{IL23R} and 5p13 associated regions for Crohn's disease (Figures
\ref{fig1}--\ref{fig3}), and lead to the accurate identification of the haplotype
backgrounds with elevated disease risk. This can aid selection of
individuals for resequencing in fine-mapping studies (data not shown)
and lead to better prediction of disease risk in un-phenotyped
individuals. A clear advantage of using Bayesian methods in our approach
is that it allows us to directly estimate the probability of two
mutations versus one mutation.

As noted above, we use a model averaging approach in which we are
interested in whether a \textit{location} is associated with the
disease. Another option would be not to carry out this model averaging
and to test each branch of the tree with its own Bayes \mbox{factor}. It would
be interesting to compare these two approaches in more detail and will
be relatively straightforward since GENECLUSTER can output probabilistic
genotype calls associated with placing a mutation on each branch of the
tree. In the new approach, we will obtain a sample of Bayes factors
rather than a single Bayes factor at each location as before. Therefore,
it is likely that we will obtain larger Bayes factors in associated
regions but the smoothness of the signal we noted above will likely
disappear. Nevertheless, in the context of fine-mapping signals, to
characterize the underlying form of an association and estimate effect
sizes, it clearly makes sense to consider each branch of the tree in its
own right as we have done.

\subsection{The~WTCCC Data}

We used the same set of filtered WTCCC data used by the main study \cite{5}.
All regions of potential association had genotypes at flanking SNPs
checked by examining the intensity cluster plots. SNPs with borderline
quality cluster plots were removed and the analysis was re-run to assess
the impact on the results.

\subsection*{Software Implementation}

Our software, called GENECLUSTER, will be made publicly available at the
time of publication from the website
\href{http://www.stats.ox.ac.uk/\textasciitilde marchini/software/gwas/gwas.html}{http://www.stats.ox.ac.uk/\textasciitilde marchini/}
\href{http://www.stats.ox.ac.uk/\textasciitilde marchini/software/gwas/gwas.html}{software/gwas/gwas.html}.

This incorporates the TREESIM method for sampling marginal genealogical
trees at a given site conditional upon a set of haplotypes.

Our other software packages, HAPGEN, IMPUTE and SNPTEST, are also
available from this website.

\subsection*{Supplementary Material}

Supplementary material to this paper is available from
\href{http://www.stats.ox.ac.uk/\textasciitilde marchini/papers/GC\_SOM.pdf}{http://www.stats.ox.ac.uk/\textasciitilde marchini/papers/}
\href{http://www.stats.ox.ac.uk/\textasciitilde marchini/papers/GC\_SOM.pdf}{GC\_SOM.pdf}.

\section*{Acknowledgments}

This paper makes use of data generated by the Wellcome Trust Case
Control Consortium. A full list of the investigators who contributed to
the generation of the data is available from \href{http://www.wtccc.org.uk}{www.wtccc.org.uk}. Funding
for the project was provided by the Wellcome Trust under award 076113.
We acknowledge support from the~Wellcome Trust, the Wolfson Foundation,
the Engineering and Physical Sciences Research Council and the US
National Institute of General Medical Sciences.

\end{document}